\documentclass[sigplan,10pt]{acmart}

\renewcommand\footnotetextcopyrightpermission[1]{}
\setcopyright{none}
\acmConference[ATC'26]{2026 ACM SIGOPS Annual Technical Conference}{November 2026}{Shatin, Hong Kong}

\usepackage{amsmath,amsthm}
\usepackage[ruled, noend, vlined]{algorithm2e}
\usepackage{listings}
\usepackage{xcolor}
\usepackage{colortbl}
\usepackage{balance}
\usepackage{graphicx}
\usepackage{subcaption}
\usepackage{wrapfig}
\usepackage[font={small}]{caption}
\usepackage{booktabs}
\usepackage[export]{adjustbox}
\usepackage{enumitem}
\usepackage{multirow}
\usepackage{siunitx}
\usepackage{pifont}
\usepackage{tikz}
\usetikzlibrary{tikzmark,calc}
\usepackage[nameinlink]{cleveref}
\usepackage[title]{appendix}
\usepackage{microtype}

\newcommand{\pytorchprofiler}{PyTorch Profiler\xspace}
\newcommand{\nsight}{NVIDIA Nsight Systems\xspace}
\newcommand{\nsys}{\texttt{nsys}\xspace}
\newcommand{\eg}{\emph{e.g.}\@\xspace}

\newcommand{\etc}{\emph{etc.}\@\xspace}

\graphicspath{{figure/plot/}}

\makeatletter
\@ifclassloaded{acmart}{%
}{%
  \usepackage{titlesec}
}
\makeatother

\newcommand{\para}[1]{\noindent\textbf{#1}\hspace{0.5em}}
\makeatletter
\@ifclassloaded{acmart}{}{%
  \renewcommand{\paragraph}[1]{\para{#1}}
}
\makeatother

\DeclareMathAlphabet{\mathcal}{OMS}{cmsy}{m}{n}

\DeclareSIUnit{\operation}{op}
\newcommand{\cmark}{\ding{51}}
\newcommand{\xmark}{\ding{55}}

\definecolor{ngray}{RGB}{102,102,102}

\newcolumntype{R}[2]{%
  >{\adjustbox{angle=#1,lap=\width-(#2)}\bgroup}%
  l%
  <{\egroup}%
}

\definecolor{bg}{rgb}{0.95,0.95,0.95}

\crefname{section}{\S}{\S\S}
\crefname{subsection}{\S}{\S\S}
\crefformat{section}{\S#2#1#3}
\crefformat{subsection}{\S#2#1#3}
\crefname{figure}{Figure}{Figures}
\Crefname{figure}{Figure}{Figures}

\newcounter{magicrownumbers}

\definecolor{bgcolor}{rgb}{0.8,0.85,0.63}

\newcommand\sysname{ARGUS\xspace}

\title{\sysname{}: Production-Scale Tracing and Performance Diagnosis for over 10,000-GPU Clusters}
\author{Jiasheng Zhou}
\affiliation{\institution{Tencent}}
\author{Longbin Zeng}
\affiliation{\institution{Tencent}}
\author{Clavis Chen}
\affiliation{\institution{Tencent}}
\author{Ruiming Lu}
\affiliation{\institution{Tencent}}
\author{Qinwei Yang}
\affiliation{\institution{Tencent}}
\author{Leyi Ye}
\affiliation{\institution{Tencent}}
\author{Ray Ying}
\affiliation{\institution{Tencent}}
\author{Key Zhang}
\affiliation{\institution{Tencent}}
\renewcommand{\authorsaddresses}{}

\begin{document}
\date{}
\begin{abstract} Large-scale LLM training requires always-on, fine-grained observability for effective performance diagnosis at scale. Coarse resource monitors alone cannot localize root causes, and fine-grained profilers incur prohibitive (5\%--30\%) overheads and massive trace volumes, making always-on deployment impractical in large production clusters.
    
We propose \sysname{}, a low-overhead, fine-grained, always-on tracing and real-time analysis system for training workloads in 10,000+ GPU-scale production clusters. \sysname{} decomposes observation along the training call hierarchy into CPU call stacks, framework semantics, and GPU kernel execution, with always-on collection under a combined overhead of less than 2\%. It builds a unified data pipeline and compresses raw kernel events by approximately 3{,}700$\times$ from 10\,MB to 2.7\,KB per rank per step. Its progressive diagnosis framework automatically isolates anomalous windows, straggler ranks, and degraded kernels through iteration-time, phase-level, and kernel-level analysis. Deployed for over six months on a 10,000+ GPU production cluster, ARGUS has supported continuous fail-slow detection and performance optimization. Our case studies further demonstrate its effectiveness across representative anomalies, including compute stragglers, link degradation, pipeline-bubble amplification, FlashAttention JIT stalls, and compute stragglers masked by communication symptoms.
\end{abstract}

\settopmatter{printfolios=true}
\maketitle

\section{Introduction}

In recent years, the rapid development of large language models (LLMS)~\cite{gpt3, palm, bloom, llama3, scalinglaw} has driven a continuous expansion of training infrastructure. Today, mainstream LLMs reach hundreds of billions to trillions of parameters~\cite{switch_transformer}. A single pre-training run typically occupies thousands to tens of thousands of GPUs for weeks or even months~\cite{megascale, llm_datacenter, byterobust}. This scale transforms performance issues from occasional incidents into systematic challenges: in a 10,000-GPU cluster, performance degradation in any single component can slow down the overall training progress, and diagnosis grows super-linearly harder with scale~\cite{reliability_ml_clusters, greyhound, straggler_whatif, holmes}.

\begin{figure}[t]
\centering
\begin{subfigure}[b]{\columnwidth}
\centering
\includegraphics[width=\textwidth]{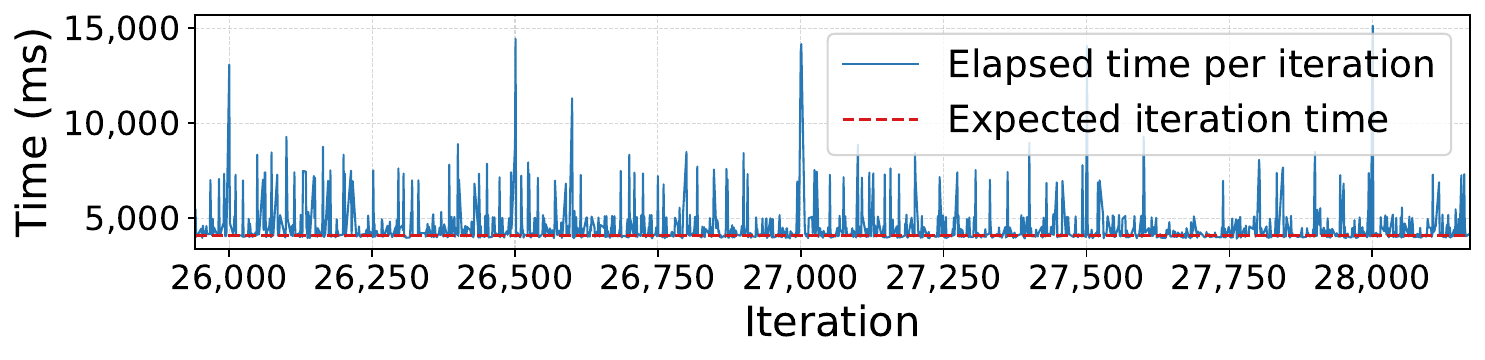}
\vspace{-0.7cm}
\caption{Iteration time spikes.}
\label{fig:iteration-spikes}
\end{subfigure}

\begin{subfigure}[b]{\columnwidth}
\centering
\includegraphics[width=\textwidth]{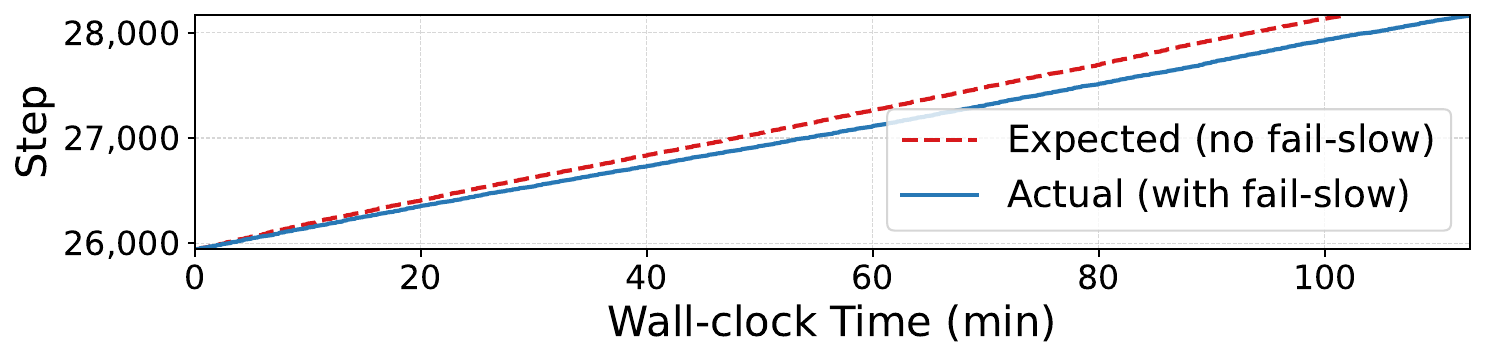}
\vspace{-0.7cm}
\caption{Cumulative progress loss.}
\label{fig:step-vs-time}
\end{subfigure}
\caption{Fail-slow in a 4096-GPU training job.}
\label{fig:iteration-time}
\end{figure}

\textbf{Performance diagnosis} for large-scale training encompasses two complementary aspects: localizing \emph{fail-slow} faults, and identifying \emph{performance optimization} opportunities. Unlike fail-stop failures that halt execution, fail-slow refers to performance degradation in any component—such as GPU hardware, communication fabric, or host-side software—that drags down the entire synchronous training job without triggering explicit errors. As shown in \cref{fig:iteration-time}, in a 4096-GPU training job, iteration time exhibits numerous spikes exceeding twice the expected duration, wasting approximately 23,758 GPU-hours (7\% of total training compute). This problem is stealthy, random, and difficult to reproduce, limiting large-scale training efficiency.

However, existing monitoring and diagnosis approaches are inadequate for effective performance diagnosis. The first class (Greyhound~\cite{greyhound}, Holmes~\cite{holmes}, C4~\cite{c4}, Minder~\cite{minder}, ByteRobust~\cite{byterobust}, Mycroft~\cite{mycroft}) adopts always-on continuous operation, tracking iteration time, communication operators, or infrastructure metrics at low overhead. They can detect anomalies and localize them to a specific machine or link, but cannot answer which kernel slowed down and why. The second class (MegaScale~\cite{megascale}, EROICA~\cite{perftracker}, FLARE~\cite{flare}) provides finer-grained traces, but each with several limitations: some remain at phase-level granularity and rely on manual post-hoc analysis; some trigger kernel-level profiling only after detecting anomalies with runtime overhead of 5\%--30\% or more, and cannot be kept continuously enabled in production training. Thus, no existing system can simultaneously achieve fine granularity, always-on operation, and real-time cross-rank analysis at 10,000-GPU scale.

Building such a system faces two core challenges. First, there is an inherent tension between fine-grained observation and low overhead. Comprehensive observation of the full training execution introduces significant runtime overhead. This not only slows training but also creates an observer effect, causing traces to capture perturbed rather than original execution behavior. Second, fine-grained trace data at 10,000-GPU scale is extraordinarily voluminous. In a 10,000-GPU cluster, each GPU generates $10^4$ to $10^5$ kernel events per minute, producing over 1\,GB/min of raw traces cluster-wide. Performing cross-rank online comparison directly on the full raw data is computationally infeasible.

To address these challenges, we propose \sysname{}, a low-overhead, fine-grained, always-on tracing and real-time analysis system for large-scale training workloads. Its design is guided by the call hierarchy of modern training systems. As shown in \cref{fig:call-hierarchy}, a training iteration spans three layers: the Python layer for scheduling and data preparation, the framework layer (\eg, Megatron~\cite{megatron, megatron_lm_sc21}, ZeRO~\cite{zero}, and FSDP~\cite{fsdp}) for phase orchestration, and the GPU runtime layer for kernel execution. Since performance anomalies may arise at any layer and manifest differently, \sysname{} decomposes observability into three corresponding signals: CPU call stacks, framework semantics, and GPU kernel traces. This decomposition enables always-on collection with a total overhead below 2\%. To support real-time analysis at 10,000-GPU scale, \sysname{} builds a unified data pipeline that streams structured metrics to a time-series database for monitoring and alerting, while persisting converted raw traces to object storage for offline analysis. For voluminous kernel traces, \sysname{} further introduces an online statistical compression method based on kernel density estimation (KDE) clustering~\cite{kde_scott}, reducing events in each time window into KB-scale summaries with a compression ratio of $10^3$--$10^4$. On top of this pipeline, \sysname{} implements a progressive diagnosis framework across different granularities, narrowing manual analysis from tens of thousands of GPUs to a few ranks and time windows. \sysname{} has been deployed in a production cluster of over 10,000 GPUs. The main contributions of this paper are as follows:
\begin{figure}[t]
\centering
\includegraphics[width=\columnwidth,page=2]{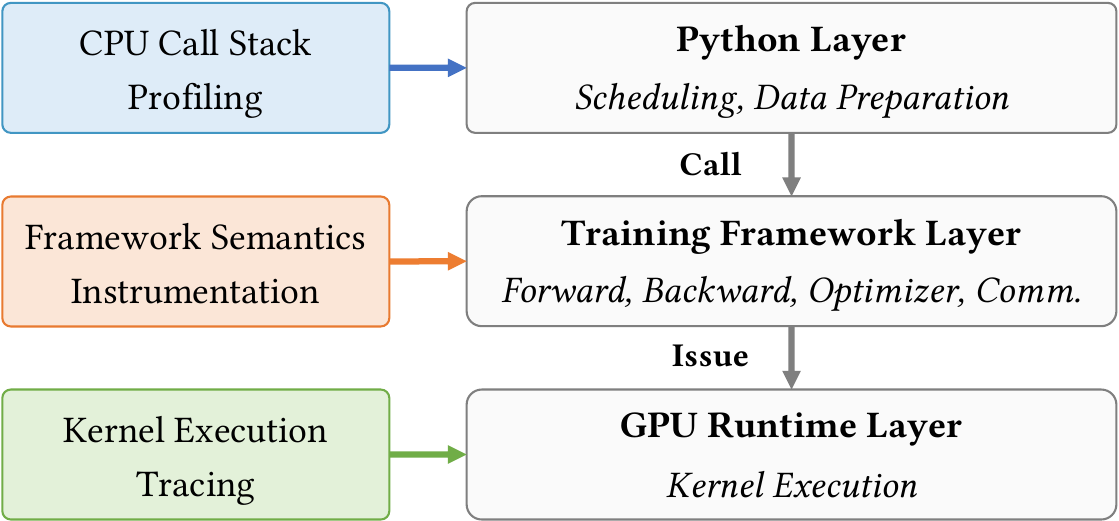}
\caption{The hierarchical structure of training execution.}
\label{fig:call-hierarchy}
\end{figure}
\begin{itemize}[leftmargin=*]
    \item We design and implement a low-overhead, fine-grained, always-on tracing system for 10,000-GPU scale clusters, decomposing observation into three independent mechanisms (CPU call stack, framework semantics, and kernel execution) that span from host-side behavior to GPU kernel execution, with a total overhead of less than 2\% (\S\ref{sec:design:monitoring}).
    \item We build a unified data pipeline that supports real-time transport, tiered storage, and online analysis. For the highest-volume kernel execution traces, we propose an online statistical compression method based on KDE clustering that achieves approximately 3{,}700$\times$ compression (from 10\,MB to 2.7\,KB per rank per step), enabling online cross-rank anomaly detection at 10,000-GPU scale (\S\ref{sec:design:pipeline}).
    \item We design a progressive diagnosis framework with parallel detection levels spanning iteration-time anomaly detection, cross-rank attribution, and kernel-level distribution comparison, narrowing the diagnostic scope from tens of thousands of ranks to single-digit suspects (\S\ref{sec:design:profiling}).
    \item We deploy \sysname{} on a production cluster of over 10,000 GPUs for more than six months, and demonstrate its practical effectiveness through five real-world case studies, diagnosing compute stragglers, communication link degradation, pipeline bubble amplification, JIT compilation blocking, and compute stragglers masked by communication symptoms (\S\ref{sec:implementation}, \S\ref{sec:evaluation}).
\end{itemize}

\section{Motivation and Design Space}

\subsection{Motivation}

The fail-slow problem in large-scale LLM training is not a simple training interruption, but rather the performance degradation of a few ranks, GPUs, links, or host-side components in synchronous training that drags down overall progress~\cite{greyhound, straggler_whatif}.
Existing resource-level monitoring can detect the presence of anomalies but cannot explain \emph{why}: it cannot distinguish whether a kernel has become slower, communication is blocked, a GPU idle gap has appeared, data loading is stalled, or Python GC is interfering.

A tracing and diagnosis system for 10,000-GPU scale clusters must simultaneously satisfy multiple interrelated requirements.
At the system-property level, GPU time is extremely expensive and any monitoring overhead directly translates into wasted compute, so the system must maintain \emph{low overhead}.
At the same time, merely identifying which machine slowed down is insufficient to guide remediation---the system needs multi-level \emph{fine granularity} observability spanning from training semantics to individual kernels.
Furthermore, fail-slow events are intermittent and unpredictable, requiring the system to operate \emph{always-on} rather than relying on manual triggering or short-window sampling.
At 10,000-GPU scale, every second of performance regression wastes substantial compute, demanding \emph{real-time analysis} capability that delivers diagnostic conclusions within minutes.
At the diagnostic-capability level, the system must automatically identify stragglers among over 10,000 ranks and progressively narrow the scope for \emph{fail-slow localization}, while also supporting parallel-strategy configuration, communication--computation overlap analysis, and operator efficiency evaluation for \emph{performance optimization}~\cite{megascale, alpa}.

\subsection{Limitations of Existing Systems}

\begin{table}[t]
\centering
\caption{Capability comparison of existing systems. \cmark{} indicates full support, $\triangle$ indicates partial support or with limitations, and \xmark{} indicates no support. Analysis timeliness: \emph{Cont.}=continuous, \emph{Trig.}=triggered after anomaly detection.}
\label{tab:comparison}
\resizebox{\columnwidth}{!}{%
\begin{tabular}{lcccccc}
\toprule
\multirow{2}{*}{\textbf{System}} & \multicolumn{4}{c}{\textbf{System Properties}} & \multicolumn{2}{c}{\textbf{Diagnostic Capabilities}} \\
\cmidrule(lr){2-5} \cmidrule(lr){6-7}
& \textbf{Low} & \textbf{Fine} & \textbf{Always-} & \textbf{Real-time} & \textbf{Fail-slow} & \textbf{Perf.} \\
& \textbf{overhead} & \textbf{granularity} & \textbf{on} & \textbf{analysis} & \textbf{localization} & \textbf{optimization} \\
\midrule
Greyhound~\cite{greyhound}     & \cmark & \xmark     & \cmark & Trig. & \cmark      & \xmark \\
Holmes~\cite{holmes}           & \cmark & \xmark     & \cmark & Trig. & \cmark      & \xmark \\
C4~\cite{c4}                   & \cmark & \xmark     & \cmark & Cont. & $\triangle$ & \xmark \\
Minder~\cite{minder}           & \cmark & \xmark     & \cmark & Cont. & $\triangle$ & \xmark \\
ByteRobust~\cite{byterobust}   & \cmark & \xmark     & \cmark & Cont. & $\triangle$ & \xmark \\
Mycroft~\cite{mycroft}         & \cmark & $\triangle$ & \cmark & Cont. & $\triangle$ & $\triangle$ \\
\midrule
MegaScale~\cite{megascale}     & \cmark & \xmark     & \cmark & Trig. & $\triangle$ & \cmark \\
EROICA~\cite{perftracker}      & $\triangle$ & \cmark & \xmark & Trig. & \cmark      & \cmark \\
FLARE~\cite{flare}             & \cmark & \cmark     & \cmark & Cont. & $\triangle$ & \cmark \\
\midrule
\textbf{\sysname}                 & \textbf{\cmark} & \textbf{\cmark} & \textbf{\cmark} & \textbf{Cont.} & \textbf{\cmark} & \textbf{\cmark} \\
\bottomrule
\end{tabular}%
}
\end{table}

As shown in Table~\ref{tab:comparison}, existing methods can be broadly categorized into two approaches, neither fully satisfying all of the above requirements.
The first class of methods (Greyhound~\cite{greyhound}, Holmes~\cite{holmes}, C4~\cite{c4}, Minder~\cite{minder}, ByteRobust~\cite{byterobust}, Mycroft~\cite{mycroft}, \etc) choose always-on continuous operation, tracking iteration time, communication operator duration, communication-layer dependencies, or infrastructure metrics at low overhead. CCL-customized systems such as Mycroft and Aegis~\cite{aegis} further improve runtime diagnosis, but remain primarily bounded by communication-layer observability. They can rapidly detect anomalies and localize them to a specific machine, link, or communication dependency, but cannot answer which arbitrary GPU kernel slowed down and why.
The second class of methods (MegaScale~\cite{megascale}, EROICA~\cite{perftracker}, FLARE~\cite{flare}, \etc) provide finer-grained traces, but each with different limitations: MegaScale remains at phase-level granularity and relies on manual post-hoc analysis; EROICA triggers kernel-level profiling only after detecting anomalies, potentially missing critical windows of sporadic events; FLARE restricts coverage to a predefined operator set with coverage bounded by explicitly instrumented operators.
More importantly, although systems like C4, Minder, and FLARE achieve continuous analysis, their diagnostic granularity remains limited to machine-level, link-level, or predefined operator-level identification. No existing system can perform kernel-level online cross-rank comparison at 10,000-GPU scale: the first class lacks the observability depth to pinpoint which specific kernel degraded, while the second class's trace data volume makes online cross-rank comparison infeasible.
In summary, existing systems either run continuously but lack sufficient diagnostic fidelity, or provide fine-grained traces but cannot be continuously deployed and scaled to 10,000+ GPUs for online cross-rank analysis---no single system satisfies all requirements in \S2.1.

\subsection{Challenges}
\label{sec:challenges}

To realize a 10,000-GPU scale tracing system that simultaneously achieves fine-grained, always-on, and real-time capabilities, two core challenges must be addressed.

\para{Challenge 1: Observer effect under a strict overhead budget.}
Accurately localizing fail-slow requires multi-level fine-grained observation of the training execution process.
General-purpose profilers and fine-grained GPU kernel profilers can achieve such coverage by collecting a wide range of information including kernel activity, CUDA API calls, memory allocations, Python call stacks, and even programmable kernel-level measurements, but at substantial runtime and data-volume cost~\cite{perftracker, neutrino}.
More critically, this overhead induces an observer effect, directly slowing training and potentially distorting the observed behavior.

\para{Challenge 2: Real-time analysis over massive distributed traces.}
Even with low-overhead collection, kernel traces at 10,000-GPU scale still produce enormous event volumes.
In a 10,000-GPU training job, each GPU generates $10^4$--$10^5$ kernel events per minute, producing 6--60\,GB/min of raw traces cluster-wide.
Performing cross-rank comparison directly on full raw data is infeasible in both computation and storage; yet over-aggregating would lose the information needed to pinpoint anomalous kernels.

These challenges show that \sysname is not enabling a lighter-weight profiler, but requires rethinking observation scope, data representation, and diagnosis workflow.

\subsection{Design Space Under Production Constraints}

The resolution of these challenges does not admit a single optimal solution, but rather involves tradeoff choices.

\para{Observation scope---coverage vs. perturbation.}
At one end of the design space lies the comprehensive view of a single profiler (strong diagnostic capability but cannot be always-on); at the other end lies low-overhead monitoring of coarse-grained metrics (can be always-on but diagnostically insufficient).
\sysname chooses a middle path: decomposing observation by training execution hierarchy, collecting CPU call stack, framework semantics, and kernel activity as complementary signals, each responsible for a single type of information with bounded overhead (detailed in \S\ref{sec:design:monitoring}).

\para{Trace representation---fidelity vs. scalability.}
Raw traces offer the highest fidelity but cannot be directly used for online transport, storage, and cross-rank analysis at 10,000-GPU scale in production; pure metrification is insufficient to localize specific kernels responsible for anomalies.
\sysname chooses tiered data representation: the online path uses structured metrics and kernel statistical summaries to support real-time queries and anomaly detection, while complete Perfetto traces are persisted to object storage for deep analysis of anomalous windows (detailed in \S\ref{sec:design:pipeline}).

\para{Diagnosis workflow---speed vs. depth.}
Performing fine-grained search across all ranks, all kernels, and all time windows is prohibitively expensive; outputting only anomalous time intervals cannot guide remediation.
\sysname chooses progressive diagnosis: multiple detection levels run in parallel, each covering a different granularity---from detecting anomalous time periods, to simultaneously localizing straggler ranks and bottleneck phases, to identifying anomalous kernels. Finally, deep-dive confirmation is performed on a small number of ranks and windows (detailed in \S\ref{sec:design:profiling}).

These three design choices jointly shape \sysname's architecture: low-overhead runtime monitoring (\S\ref{sec:design:monitoring}), scalable trace processing (\S\ref{sec:design:pipeline}), and progressive diagnosis (\S\ref{sec:design:profiling}).

\section{System Overview}
\label{sec:design}

\subsection{Architecture and Data Flow}
\label{sec:design:overview}

\begin{figure}[t]
\centering
\includegraphics[width=\columnwidth,page=1]{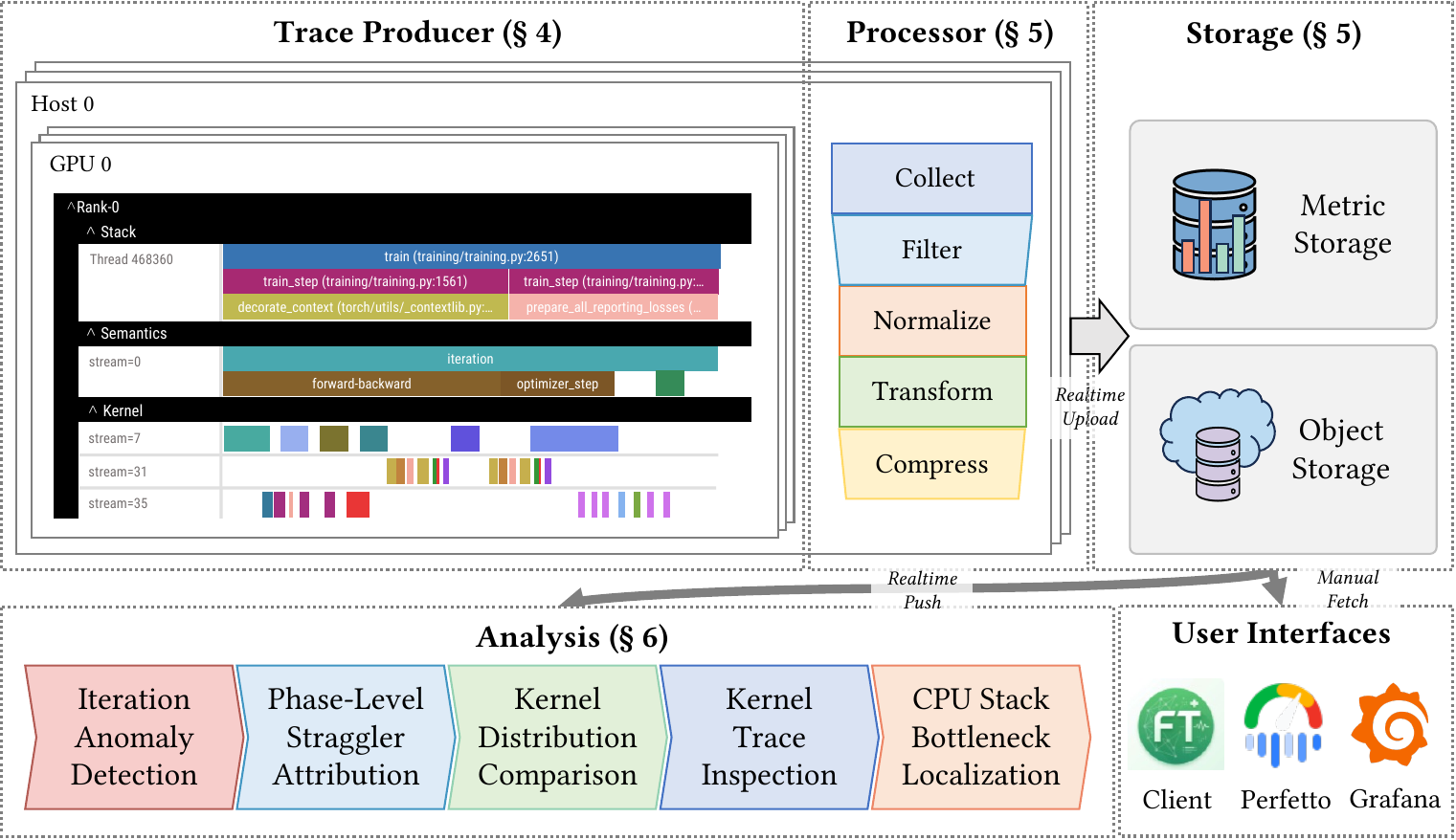}
\caption{Overall architecture of \sysname.}
\label{fig:arch}
\end{figure}

\noindent\textbf{Trace Producer (\S\ref{sec:design:monitoring}).}
Deployed within each training process, continuously producing three complementary observations with low overhead.
CPU call stack profiling captures Python call stacks via external sampling; framework semantics instrumentation records the GPU-side duration of each training phase; kernel execution tracing continuously records every kernel's launch time, duration, and stream.

\noindent\textbf{Processor (\S\ref{sec:design:pipeline}).}
An independent process deployed per host, receiving raw event streams from all local Trace Producers. It filters, normalizes, and converts events into Perfetto~\cite{perfetto} format, and performs online statistical compression on the kernel trace, condensing kernel events within each time window into KB-scale structured statistical summaries.

\noindent\textbf{Storage (\S\ref{sec:design:pipeline}).}
Tiered storage. Metric Storage ingests structured metrics and kernel statistical summaries, supporting Grafana~\cite{grafana} visualization and low-latency alerting in real time. Object Storage persists complete Perfetto trace files.

\noindent\textbf{Analysis (\S\ref{sec:design:profiling}).}
Implements automated detection algorithms that identify anomalous windows, straggler ranks, and degraded kernels, and determines the scope for confirmation.

\subsection{User Interfaces}
\label{sec:design:interfaces}

\sysname's user-facing entry point is FT-Client, a unified diagnostic interface integrating real-time monitoring and deep analysis capabilities. After specifying a training job and time range, FT-Client presents results through two visualization systems: the Grafana Dashboard displays per-rank iteration time, phase durations, kernel anomaly alerts, and cross-rank comparisons in real-time. Perfetto loads execution traces from object storage for deep-dive analysis, presenting kernel execution timing, semantic phase durations, and CPU call stacks in a unified timeline view.

\section{Low-overhead Runtime Monitoring}
\label{sec:design:monitoring}

This section instantiates the observation-scope choice from \S2.4, decomposing observation into three hierarchy-specific signals. \sysname does not use a single profiling tool to comprehensively capture all information; instead, each signal targets a specific layer of the execution hierarchy. The three collection channels are complementary rather than substitutive; starting or stopping any one does not affect the others.

\subsection{CPU Stack Sampling}
\label{sec:design:pyspy}

Fail-slow causes do not always originate from the GPU. Python GC pauses, data loading stalls, CPU contention, and GIL contention can equally lead to iteration time anomalies. \sysname employs py-spy~\cite{pyspy} to obtain Python call stacks by reading the target process's memory, requiring no modifications to training code or injection of hooks. We adapt it for streaming: instead of generating a single flamegraph, it continuously outputs structured call stack snapshots in fixed sampling windows, giving CPU-side observation the same temporal continuity as GPU-side traces. When semantics shows an anomalously long phase but kernel execution tracing indicates normal GPU execution time, the CPU call stack from the same window can quickly identify host-side stalls (such as GIL contention, data preprocessing bottlenecks, or anomalous system calls), avoiding misattribution of non-GPU problems to kernel or communication issues.

\subsection{Framework Semantics}
\label{sec:design:instrumentation}

\sysname inserts CUDA Events at the entry and exit of key framework phases (forward, backward, optimizer, communication), using the elapsed time between two events on the same CUDA stream to capture the true GPU-side execution duration of that semantic interval~\cite{cuda_events}. Unlike CPU-side wall-clock time, a CUDA Event is timestamped only when the GPU actually reaches that position in the stream's execution order, thus more accurately reflecting device-side execution time and being less affected by CPU scheduling, asynchronous submission, and host/device decoupling. The instrumentation does not modify the framework's internal implementation or optimizer logic; it performs lightweight wrapping only at call sites where semantic boundaries are clear, concentrating profiling logic on a few critical paths.

A key challenge is accurately identifying the CUDA stream on which the target phase actually executes. Computation operations typically run on the default stream, but communication operations use NCCL~\cite{nccl} internal streams. If events are inserted on the default stream, communication duration is near-zero. Therefore, the actual execution stream must be determined based on communication type: collective communication selects the stream based on device, P2P communication based on peer group rank~\cite{pytorch}. 

\subsection{Kernel Activity}
\label{sec:design:cupti}

\sysname continuously records GPU kernel names, launch times, durations, and stream assignments via the CUPTI Activity API~\cite{cupti}, injecting into the training process via environment variables without requiring modifications to the training framework or user code. To avoid the callback path becoming a bottleneck on the training hot path, \sysname organizes the tracing backend into three decoupled paths: the \emph{control path} only transmits start/stop signals; the \emph{collection path} performs only the most lightweight operation in the callback---receiving the buffer and handing it off to the backend via a queue; the \emph{processing and export path} asynchronously completes parsing, format conversion, and disk writes in an independent thread without blocking the frontend collection. Beyond this architecture design, \sysname further controls overhead and stability through selective process injection, pre-allocated buffer reuse, and bounded resources with backpressure (detailed in Appendix~\ref{sec:appendix:cupti}).

\vspace{-0.3cm}
\section{Scalable Trace Processing}
\label{sec:design:pipeline}

This section instantiates the trace-representation choice from \S2.4, transforming heterogeneous traces into scalable online summaries and complete persisted traces.

\subsection{Data Pipeline Architecture}
\label{sec:design:pipeline:arch}

\begin{figure}[t]
\centering
\includegraphics[width=\columnwidth,page=4]{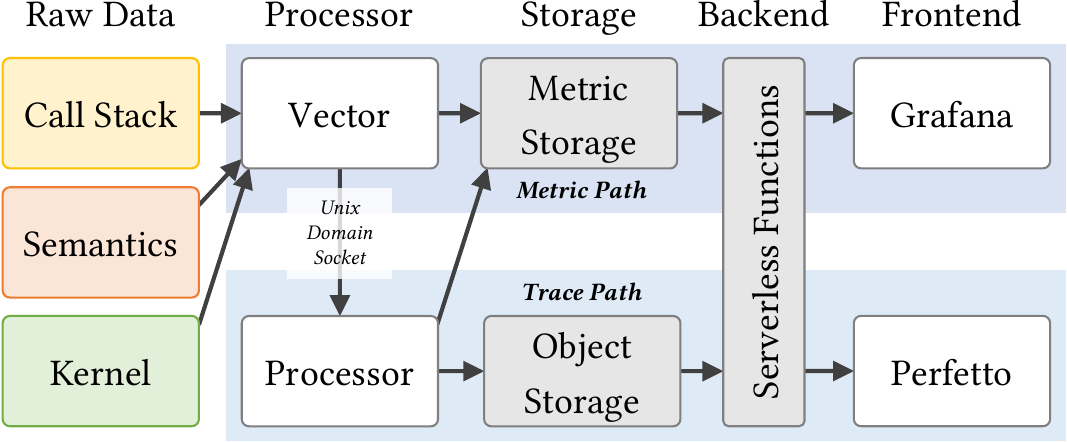}
\caption{Data pipeline architecture. 
}
\label{fig:pipeline}
\end{figure}

\sysname interposes a unified data pipeline between runtime monitoring and progressive diagnosis. The ingestion and transformation layer is implemented using Vector~\cite{vector} for its high-throughput and lightweight transformation capabilities. As shown in Figure~\ref{fig:pipeline}, Vector ingests three types of local observation logs and distributes them to two downstream paths based on their characteristics. The \textbf{trace path} forwards raw event streams via Unix domain socket to an independent Processor process. The Processor is responsible for two tasks. First, it filters, normalizes, and converts raw events into Perfetto format, writing them to Object Storage for subsequent deep analysis. Second, it performs online statistical compression on kernel traces (\S\ref{sec:design:compression}), writing compressed structured summaries to Metric Storage for real-time cross-rank comparison. The \textbf{metrics path} handles directly quantifiable observation results (phase duration, iteration time), writing them to Metric Storage via the Prometheus Remote Write protocol~\cite{prometheus} to support real-time Grafana visualization and low-latency alerting. 

This tiered design separates metrics for real-time monitoring from traces that require preservation of complete temporal semantics, and decouples lightweight data ingestion from the heavier trace materialization process, ensuring that neither process becomes a bottleneck for the other.

\begin{figure}[t]
\centering
\includegraphics[width=\columnwidth,page=3]{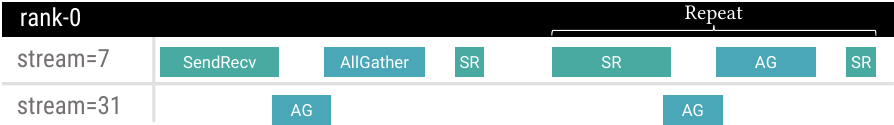}
\caption{Repetitive kernel execution patterns.} 
\label{fig:pattern}
\end{figure}

\subsection{Online Statistical Compression of Kernel Traces}
\label{sec:design:compression}

In a 10,000-GPU cluster, each GPU can produce $10^4$--$10^5$ kernel events per minute, and the total raw trace volume per hour can reach hundreds of gigabytes. Performing cross-rank anomaly analysis directly on the full raw data is infeasible in both computation and storage. To address this, \sysname designs an online statistical compression method within the Processor, transforming raw kernel events into compact structured summaries that enable subsequent cross-rank anomaly detection (\S\ref{sec:design:kernel_detection}) on KB-scale data.

\noindent\textbf{Rationale for statistical compression.}
As shown in Figure~\ref{fig:pattern}, kernel execution in distributed training exhibits highly regular repetitive patterns: each stream on the same rank repeatedly executes the same combination of kernels in a fixed sequence (\eg, stream=7 exhibits a SendRecv--AllGather--SendRecv alternating sequence), and the kernel durations at the same position across repetitions is highly consistent. Furthermore, ranks with the same parallel role execute identical kernel sequences, so under normal conditions the duration distribution of the same kernel across ranks should exhibit highly consistent statistical characteristics. This regularity makes it possible to extract statistics for each kernel and perform cross-rank comparison.

However, directly computing a single statistic across all duration samples of the same kernel is problematic. Although kernels of the same name at the same position have highly consistent durations, kernels of the same name at different positions may differ significantly: as shown in Figure~\ref{fig:pattern} for stream=7, the first SendRecv and the third SendRecv on the same stream differ in duration by several multiples due to different data transfer volumes; AllGather kernels on different streams (\eg, stream=7 vs. stream=31) differ even more dramatically in time scale due to participation in different communication groups. This multimodal structure means that computing a global median without distinction would mask positional and stream-level differences, leading to false positives in anomaly detection. Therefore, \sysname first clusters kernel durations to identify each mode, and then extracts statistics separately for each mode.

\noindent\textbf{Clustering method.}
The online clustering algorithm must satisfy three constraints. First, it must not require pre-specifying the number of clusters $K$, since the number of modes varies greatly across different kernels. Second, it must not require historical data or a warm-up phase, as the system should be able to run independently on any time window. Third, its time complexity must be linear in the number of samples, as online deployment requires computation to complete by the end of each window. Based on these requirements, \sysname adopts the valley detection method based on kernel density estimation (KDE): it constructs a density curve over log-duration samples and uses local minima (valleys) as cluster boundaries, with the number of clusters automatically determined by the data's modal structure.

\begin{figure}[t]
\centering
\begin{subfigure}[b]{0.48\columnwidth}
\centering
\includegraphics[width=\textwidth]{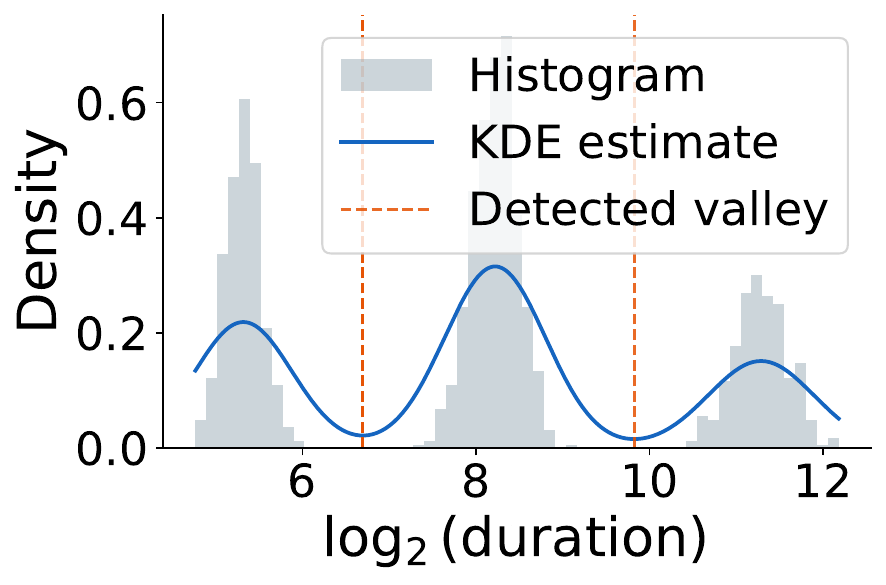}
\caption{KDE valley detection.}
\label{fig:cluster-kde}
\end{subfigure}%
\hfill%
\begin{subfigure}[b]{0.48\columnwidth}
\centering
\includegraphics[width=\textwidth]{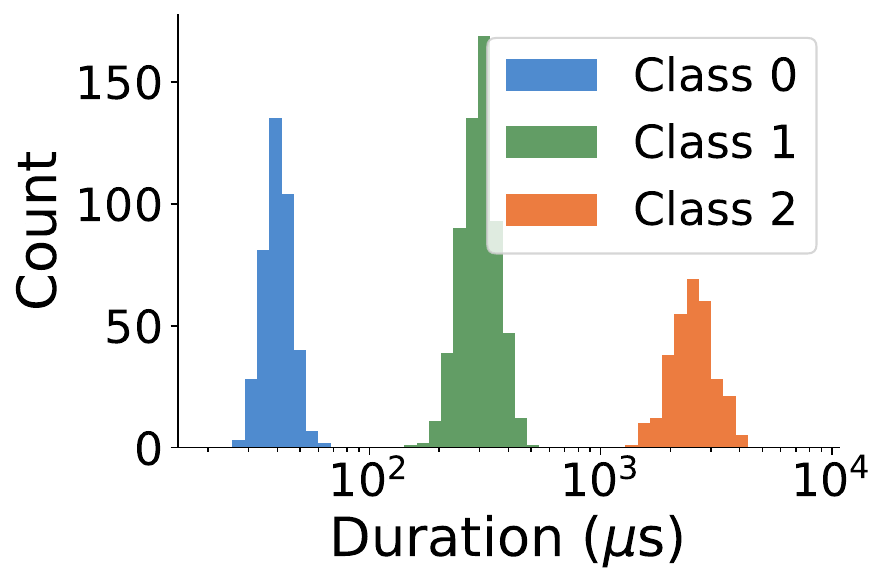}
\caption{Resulting clusters.}
\label{fig:cluster-linear}
\end{subfigure}
\caption{Visualization of KDE-based clustering. 
}
\label{fig:cluster}
\end{figure}

\noindent\textbf{Algorithm procedure.}
For each $(kernel, stream, rank)$ combination in each time window, the algorithm first applies a logarithmic transformation to the raw durations, then computes the KDE density function on an equally-spaced grid for subsequent valley detection:
\begin{equation}
\setlength{\abovedisplayskip}{1pt}
\setlength{\belowdisplayskip}{1pt}
\hat{f}(x) = \frac{1}{n} \sum_{i=1}^{n} K\!\left(\frac{x - x_i}{h}\right)
\end{equation}
\noindent where $K(\cdot)$ is the Gaussian kernel function~\cite{kde_scott} and the bandwidth $h$ is automatically determined by Scott's rule~\cite{kde_scott}: $h = 1.06 \cdot \sigma \cdot n^{-1/5}$. All local minima on the density curve are identified as candidate cluster boundaries, and two layers of filtering are applied to eliminate noise: \emph{cluster-level filtering} requires that both sides of each valley contain sufficient samples, preventing noise fluctuations from being misidentified as independent modes; \emph{spacing filtering} requires that the duration difference between adjacent retained boundaries is sufficiently significant, preventing pseudo-valleys within the same peak from causing over-segmentation. Figure~\ref{fig:cluster} visualizes this process. In log-space, the KDE density curve clearly exhibits multi-peak structure, and the algorithm determines cluster boundaries at density minima (panel~(a)). Mapping these boundaries back to linear duration space partitions the original samples into discrete clusters, each corresponding to a characteristic execution time scale (panel~(b)). Finally, three statistics are computed for each cluster: execution count $count$, median duration $p50$, and 99th percentile duration $p99$.

\noindent\textbf{Compression effectiveness.}
Through the above method, all raw events for each $(kernel, stream, rank)$ within a time window are compressed into a few cluster triples $(count, p50, p99)$. Taking 10,000-GPU scale as an example, if each rank has approximately 100 active $(kernel, stream)$ combinations with an average of 2 clusters each, then each rank needs to upload only about 200 statistical records per minute ($\sim$several KB), and the total volume from 10,000 ranks is approximately tens of MB; compared to the hundreds of GB of raw traces, this achieves a compression ratio of $10^3$--$10^4$. This makes real-time comparative analysis across all ranks feasible in both computation and storage.

The compression is lossy but fidelity-preserving: $p50$ captures the typical execution time of each kernel class, $p99$ retains tail latency, and $count$ records the frequency proportion of each cluster. As described in \S\ref{sec:design:kernel_detection}, these three statistics are sufficient to support cross-rank anomaly detection via CDF reconstruction and Wasserstein distance~\cite{wasserstein}. The entire statistical compression process is performed within the Processor, running asynchronously from the training process without blocking the training main loop.

\section{Progressive Diagnosis}
\label{sec:design:profiling}

Building on the observation-scope and trace-representation choices from \S2.4, this section instantiates the diagnosis-workflow choice, progressively narrowing the search space.

\begin{table}[t]
\centering
\caption{Overview of \sysname diagnostic levels.}
\label{tab:levels}
\resizebox{\columnwidth}{!}{%
\begin{tabular}{@{}clcll@{}}
\toprule
Level & Data Source & Mode & Purpose & Latency \\
\midrule
L1 & Iteration time series & Auto & Anomalous time window classification & Seconds \\
L2 & Semantic phase durations & Auto & Straggler rank and bottleneck phase & Seconds \\
L3 & Kernel statistical summaries & Auto & Degraded kernel identification & Minutes \\
L4 & Execution trace & Manual & Critical path and root cause confirmation & On-demand \\
L5 & CPU call stacks & Manual & Host-side stall localization & On-demand \\
\bottomrule
\end{tabular}%
}
\end{table}

\sysname's diagnostic framework consists of five levels, with L1, L2, and L3 running in parallel as three automated levels covering different granularities (iteration time / semantic phase / kernel statistics), jointly narrowing the scope requiring manual analysis. L1 serves fail-slow detection by identifying anomalous time windows. L2 and L3 serve both fail-slow localization and performance optimization: they pinpoint straggler ranks and degraded kernels, while also revealing inefficiencies such as load imbalance or suboptimal communication overlap. L4/L5 provide high-fidelity deep-dive confirmation primarily for performance optimization, enabling engineers to inspect execution traces for root-cause analysis and optimization opportunities, as well as offline critical path analysis and host-side idle-cause localization. At 10,000-GPU scale, performing fine-grained analysis on all ranks is neither economical nor necessary; the three automated levels reduce the search scope from tens of thousands of ranks to single-digit ranks and time windows.

\subsection{Iteration-level Detection (L1) and Phase-level Attribution (L2)}
\label{sec:design:iter_detection}

L1 continuously collects each rank's iteration time series, running two complementary anomaly detection algorithms: sliding-window ratio-gated jitter detection for short-term fluctuations and spikes, while full-scan change-point detection for step-wise regression. Together they classify iteration time behavior as stable, jitter, regression, or both.

L2 performs cross-rank comparison on the phase durations recorded by framework semantics instrumentation within parallelism groups, using the coefficient of variation (CV) to quantify intra-group inconsistency and z-scores to identify straggler ranks. Its key design is \textbf{parallelism-group-aware routing}: since distributed training uses multiple parallelism dimensions simultaneously~\cite{megatron, megatron_lm_sc21, zero, fsdp, alpa}, each event must be compared only among ranks that share the same parallel role. The system maintains a routing table that maps each semantics event to its corresponding parallelism group, as illustrated in Table~\ref{tab:routing}. For pure computation events such as \texttt{self\_attention} and \texttt{moe\_experts}, a high CV within the corresponding group directly indicates a straggler rank whose compute is slower. For communication events such as \texttt{dp-allreduce} and \texttt{ep-alltoall}, the system additionally determines whether the prolonged duration originates from the rank itself or waiting for a slow peer in the same synchronization group. The complete algorithm descriptions for L1 and L2 are provided in Appendix~\ref{sec:appendix:diagnosis}.

\begin{table}[t]
\centering
\caption{Representative parallelism-group-aware routing rules.}
\label{tab:routing}
\resizebox{0.9\columnwidth}{!}{%
\begin{tabular}{@{}lc@{}}
\toprule
Event Type & Comparison Group \\
\midrule
\texttt{gated\_mla\_self\_attention}, \texttt{self\_attention} & DP group \\
\texttt{moe\_layer}, \texttt{moe\_experts} & EP group \\
\texttt{dp-allreduce}, \texttt{dp-reduce-scatter} & DP group \\
\texttt{ep-alltoall} & EP group \\
\ldots & \ldots \\
\bottomrule
\end{tabular}%
}
\end{table}

\subsection{Kernel Statistics Anomaly Detection (L3)}
\label{sec:design:kernel_detection}

\begin{figure}[t]
\centering
\begin{subfigure}[b]{0.32\columnwidth}
\centering
\includegraphics[width=\textwidth]{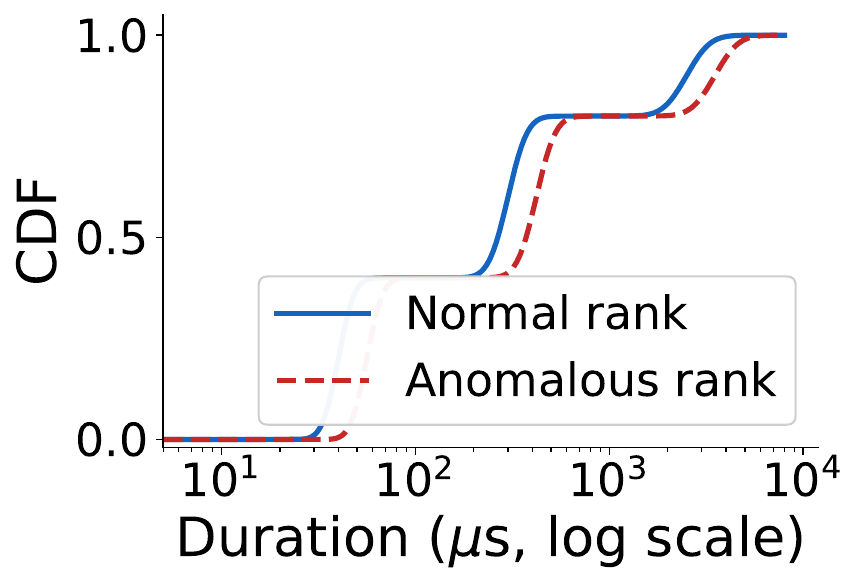}
\caption{CDF reconstruction.}
\label{fig:detection-cdf}
\end{subfigure}%
\hfill%
\begin{subfigure}[b]{0.32\columnwidth}
\centering
\includegraphics[width=\textwidth]{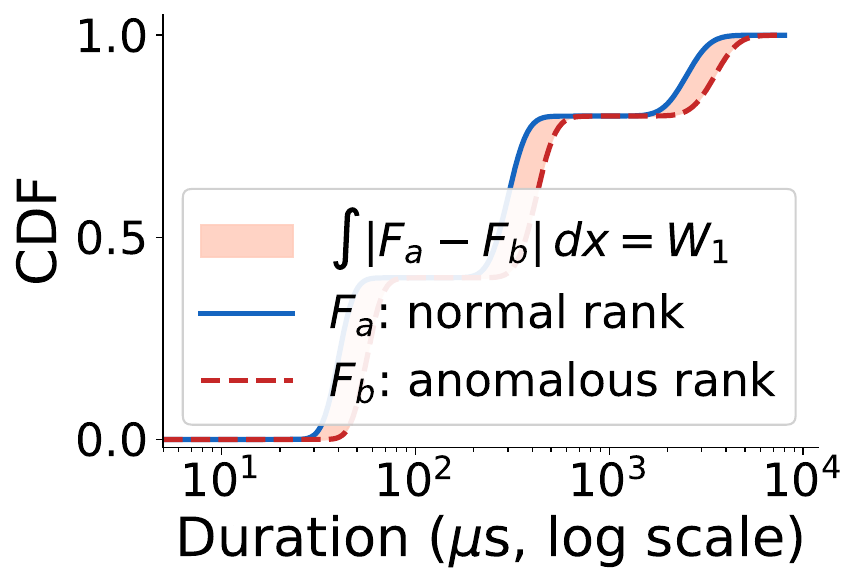}
\caption{$W_1$ distance.}
\label{fig:detection-w1}
\end{subfigure}%
\hfill%
\begin{subfigure}[b]{0.32\columnwidth}
\centering
\includegraphics[width=\textwidth]{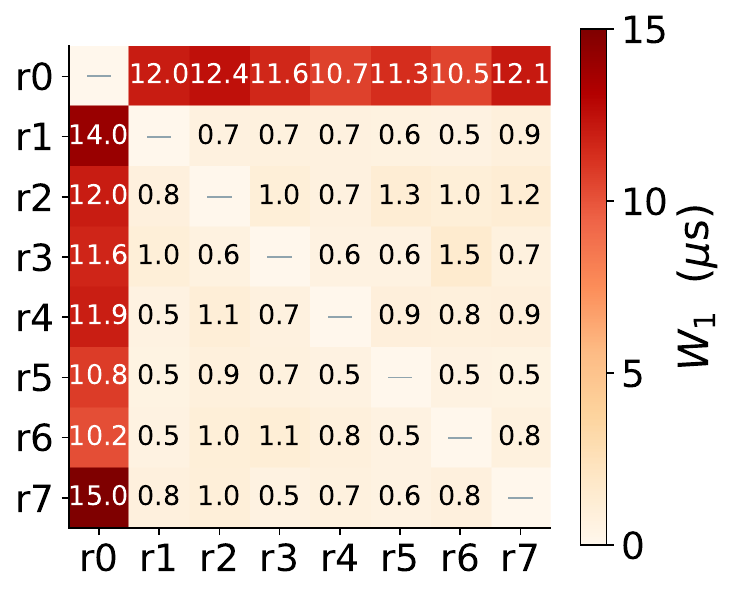}
\caption{Distance matrix.}
\label{fig:detection-matrix}
\end{subfigure}
\caption{Kernel statistics anomaly detection workflow. 
}
\label{fig:detection}
\end{figure}

Based on the compressed statistics described in \S\ref{sec:design:compression}, \sysname designs a cross-rank anomaly detection method that identifies performance anomalies by comparing the execution time distribution of the same kernel across different ranks. The core idea is that in synchronous distributed training, ranks with the same parallel role should execute identical kernel sequences, and their duration distributions should be highly consistent. When the distribution of a specific kernel on a particular rank significantly deviates from other ranks, that rank is identified as a fail-slow suspect. Unlike the phase-duration-based cross-rank comparison in \S\ref{sec:design:iter_detection}, this level of analysis drills down to individual kernel granularity, precisely identifying which specific kernel is behaving anomalously, and providing a precise entry point for subsequent root-cause confirmation via Perfetto traces.

Figure~\ref{fig:detection} illustrates the detection workflow: (a)~CDFs are reconstructed from compressed statistics for each rank, where the anomalous rank's CDF curve clearly deviates from normal ranks; (b)~the Wasserstein-1 distance between any two ranks' reconstructed CDFs quantifies distribution difference; (c)~the $W_1$ values for all rank pairs form a matrix, where the anomalous rank shows systematically elevated distances to others, enabling accurate identification.

\noindent\textbf{CDF reconstruction.}
At 10,000-GPU scale, directly performing cross-rank comparison analysis on full raw samples is infeasible; therefore, \sysname reconstructs cumulative distribution functions (CDFs) from the compressed statistics obtained in \S\ref{sec:design:compression} for distribution comparison. \sysname employs a parametric reconstruction method based on the log-normal assumption: for each cluster $c$, a log-normal component is constructed with location parameter $\mu_c = \ln(p50_c)$, positioning the distribution center at the median, and scale parameter $\sigma_c = \frac{\ln(p99_c) - \ln(p50_c)}{2.326}$, fitting the tail shape using the relationship with the standard normal distribution's 99th percentile point $z_{0.99} = 2.326$. The components are weighted by $count$ to form a mixture CDF:
\begin{equation}
\setlength{\abovedisplayskip}{1pt}
\setlength{\belowdisplayskip}{1pt}
F(x) = \sum_{c} \frac{count_c}{\sum_{c'} count_{c'}} \cdot \Phi\!\left(\frac{\ln x - \mu_c}{\sigma_c}\right)
\end{equation}
\noindent where $\Phi(\cdot)$ is the CDF of the standard normal distribution. This reconstruction method utilizes median and tail information, capturing both the overall shape and tail latency characteristics of the distribution. The log-normal assumption is chosen because kernel durations in practice exhibit a clearly right-skewed distribution that approximates normality after log-transformation, allowing each mode's shape to be well parameterized using only two percentiles (p50, p99).

\noindent\textbf{Wasserstein-1 distance and anomaly identification.}
From the reconstructed CDFs, \sysname uses the Wasserstein-1 distance ($W_1$, also known as Earth Mover's Distance) to quantify distribution differences between two ranks:
\begin{equation}
\setlength{\abovedisplayskip}{1pt}
\setlength{\belowdisplayskip}{1pt}
W_1(F_a, F_b) = \int_0^{\infty} |F_a(x) - F_b(x)| \, dx
\end{equation}
This distance is computed via trapezoidal numerical integration on a log-uniform grid. The choice of $W_1$ over KL divergence or the KS statistic is motivated by the following: $W_1$ possesses true metric properties (satisfying the triangle inequality), is sensitive to both shifts and scaling of distributions, and has an intuitive physical interpretation---it can be understood as the minimum work required to ``transport'' one distribution into another~\cite{wasserstein}. For fail-slow detection scenarios, $W_1$ can simultaneously capture overall distribution shift (persistent slowdown) and tail inflation (intermittent tail latency), whereas the KS statistic only considers the maximum deviation point, and KL divergence is unstable when distribution supports do not completely overlap.

For each $(kernel, stream)$ combination, the system computes the $W_1$ values between all pairs of ranks and organizes them into a distance matrix. Under normal conditions, ranks with the same parallel role execute identical kernel sequences, and pairwise $W_1$ values should be relatively small. If a rank exhibits a performance anomaly, its $W_1$ with all other ranks will be systematically elevated, manifesting as significantly higher values in the corresponding row and column of the distance matrix compared to other positions.

\noindent\textbf{Cross-rank IQR anomaly determination.}
To automatically identify anomalous ranks from the distance matrix, \sysname computes the mean $W_1$ of each rank to all other ranks as that rank's deviation score, then applies a robust statistical method based on the interquartile range (IQR) for aoutlier-based nomaly determination~\cite{tukey_eda}:
\begin{equation}
\setlength{\abovedisplayskip}{1pt}
\setlength{\belowdisplayskip}{1pt}
\text{upper\_fence} = Q_3 + \alpha \cdot (Q_3 - Q_1)
\end{equation}
\noindent where $Q_1$ and $Q_3$ are the 25th and 75th percentiles of all ranks' deviation scores, respectively, and $\alpha$ is the anomaly coefficient. Ranks exceeding the upper fence are flagged as anomalous. The IQR method is chosen over mean-and-standard-deviation-based determination because IQR is robust to extreme values: even if some ranks exhibit extreme deviations, the estimates of $Q_1$ and $Q_3$ are not distorted, thus not affecting the baseline for judging other ranks.

\subsection{Deep-dive Confirmation (L4/L5)}
\label{sec:design:deepdive}

L4/L5 do not perform 10,000-GPU-wide searches; instead, they provide high-fidelity root-cause confirmation for the small number of ranks and anomalous windows identified by L1--L3. Grafana duration scatter plots reveal how a specific kernel's duration evolves over time, and heatmaps display anomaly distribution across a rank-by-time matrix. For finer-grained inspection, engineers examine the Perfetto timeline to inspect kernel execution timing, semantic phase durations, and CPU call stacks in a unified view. Beyond interactive inspection, L4 also supports offline critical path analysis using approaches similar to Holistic Trace Analysis~\cite{hta}, identifying the longest sequential dependency chain that determines iteration time. L5 performs offline analysis of CPU call stacks to localize host-side causes when both compute and communication are simultaneously idle, pinpointing which function is the contributor to the stall.

\section{Implementation}
\label{sec:implementation}

We implement the runtime monitoring components of \sysname in approximately 2.7K lines of C++, 16.7K lines of Python, and a streaming adaptation of \texttt{py-spy}.
The kernel execution tracing component is compiled as a standalone shared library (\texttt{libcupti\_injector.so}) and injected into the target training process via  \texttt{CUDA\_INJECTION\allowbreak 64\_PATH} environment variable. The CUDA runtime automatically loads this library during initialization, requiring no modifications to the training code or launch scripts.
Framework semantics instrumentation is provided as a Python package. The training framework enables it by calling the exposed API at critical path boundaries, while the underlying CUDA Event management and NCCL stream queries are implemented as C++ extensions.
The Processor is implemented in Go (approximately 7.3K lines) and receives raw traces over Unix domain sockets, performing Perfetto encoding, online kernel statistical compression, and storage writes.
The diagnosis and analysis service is implemented in Python (approximately 24K lines) and encapsulates the three-level automated detection algorithms described in \cref{sec:design:profiling}.

All components are deployed as sidecars alongside training tasks and enabled via a single environment variable, supporting batch onboarding across 10,000-GPU scale clusters.

\section{Evaluation}
\label{sec:evaluation}

This section evaluates \sysname along two dimensions: runtime overhead (\cref{sec:eval-overhead}) and data volume and compression characteristics at scale (\cref{sec:eval-compression}). 
As a complementary evaluation, fault diagnosis capability is evaluated in Appendix~\ref{sec:appendix:diagnosis-capability}.

\subsection{Experimental Setup}
\label{sec:eval-setup}

Experiments are conducted on an 8-GPU node with intra-node GPUs interconnected via NVLink. We train the HunYuan-V3 Preview model~\cite{hunyuan} (configuration details in Appendix~\ref{sec:appendix:setup}). We compare \sysname against two widely-used profiling tools:
(1)~\textbf{\pytorchprofiler}, the built-in performance analysis tool in PyTorch~\cite{kineto}, and
(2)~\textbf{\nsight}, a system-level profiling tool.
All tools are configured in always-on mode throughout training.

\subsection{Runtime Overhead}
\label{sec:eval-overhead}

\begin{figure}[t]
\centering
\captionsetup{skip=2pt}
\includegraphics[width=\columnwidth]{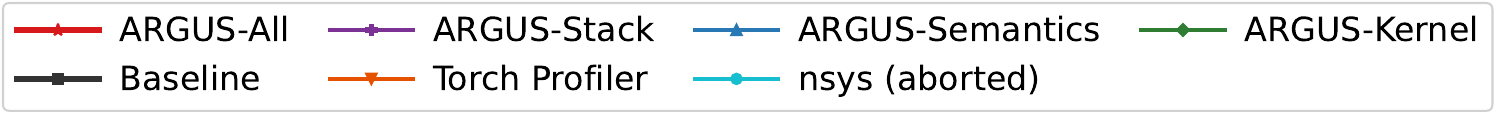}
\vspace{2pt}
\begin{subfigure}[t]{0.49\columnwidth}
  \includegraphics[width=\linewidth]{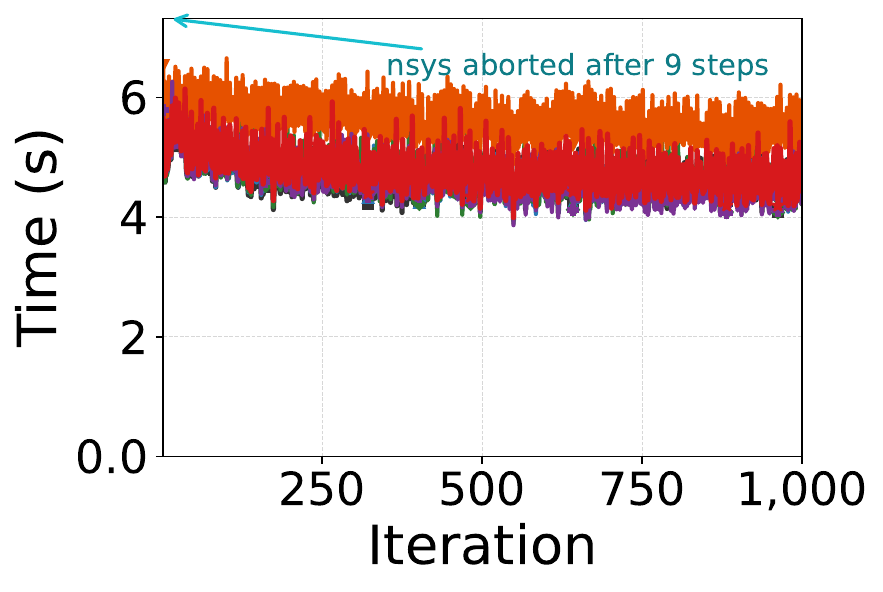}
  \caption{8-GPU}
  \label{fig:itertime-8gpu}
\end{subfigure}\hfil
\begin{subfigure}[t]{0.49\columnwidth}
  \includegraphics[width=\linewidth]{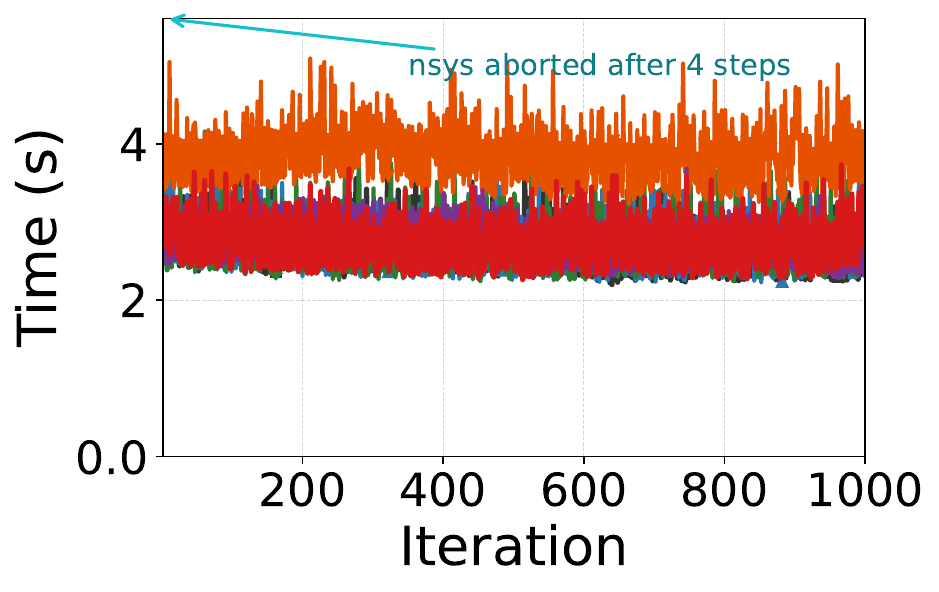}
  \caption{32-GPU}
  \label{fig:itertime-32gpu}
\end{subfigure}
\caption{Training time under different profiling configurations.}
\label{fig:itertime}
\end{figure}

\begin{figure}[t]
\centering
\captionsetup{skip=2pt}
\includegraphics[width=\columnwidth]{eval_overhead_legend.pdf}
\vspace{2pt}
\begin{subfigure}[t]{0.49\columnwidth}
  \includegraphics[width=\linewidth]{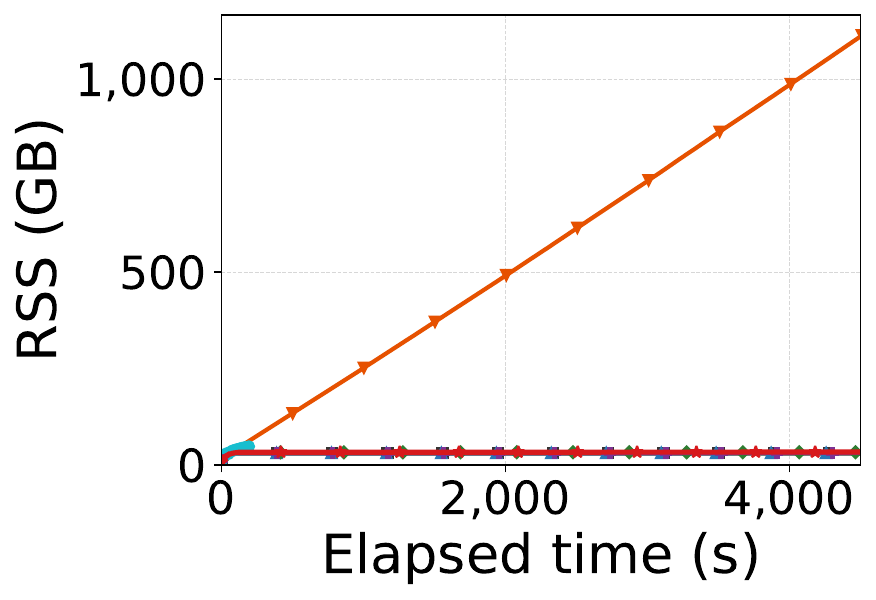}
  \caption{8-GPU}
  \label{fig:rss-8gpu}
\end{subfigure}\hfil
\begin{subfigure}[t]{0.49\columnwidth}
  \includegraphics[width=\linewidth]{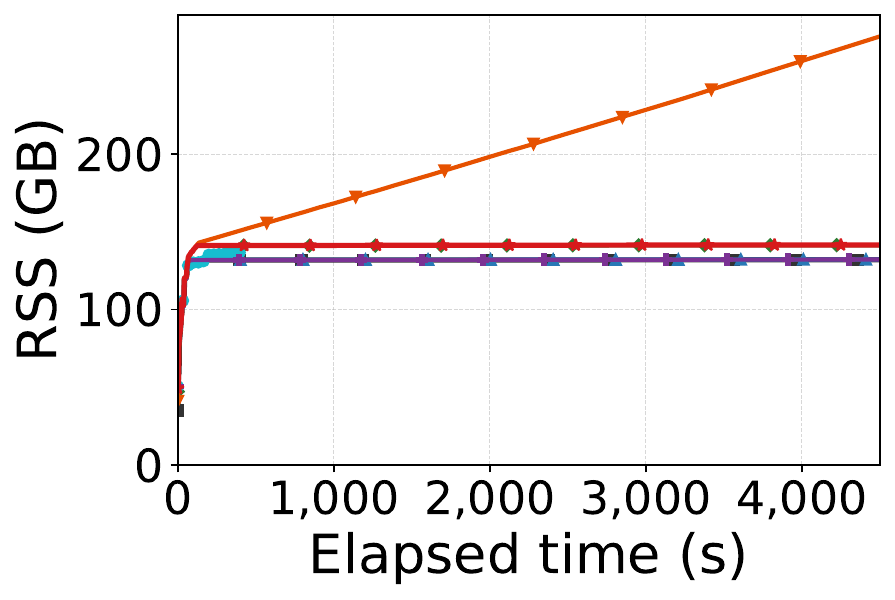}
  \caption{32-GPU}
  \label{fig:rss-32gpu}
\end{subfigure}
\caption{Resident Set Size (RSS) over time.}
\label{fig:rss}
\end{figure}

\para{Iteration time.}
As shown in \cref{fig:itertime}, we train for 1{,}000 iterations at both 8-GPU and 32-GPU scales and compare per-iteration time across configurations.
Among \sysname components, semantics instrumentation and stack sampling introduce negligible overhead, kernel execution tracing (CUPTI) adds approximately 1\%--2\%, and all three combined remain within 2\%.
As model size grows and GPU computation dominates, this overhead further diminishes.
In contrast, \pytorchprofiler inflates iteration time by 20\%--44\% and eventually triggers out-of-memory failures due to unbounded trace accumulation, making it entirely impractical for production use. \nsys fails to complete training under always-on mode at either scale: at 8-GPU, training produces NaN at iteration~10; at 32-GPU, training hangs during AllToAll communication. Neither failure occurs in the baseline, confirming the observer-effect challenge discussed in \cref{sec:challenges}: general-purpose profilers can distort or break training execution.

\para{Memory footprint.}
As shown in \cref{fig:rss}, we continuously monitor RSS across all configurations.
\sysname adds approximately 2\,GB (8-GPU) and 10\,GB (32-GPU) over the baseline, primarily from the pre-allocated CUPTI ring buffer; semantics instrumentation and stack sampling consume negligible additional memory.
Because \sysname employs a streaming architecture that hands off collected data to the Processor immediately without accumulating raw traces locally, its memory footprint remains constant regardless of training duration.
In contrast, \pytorchprofiler accumulates complete trace data in memory until the profiling window ends, causing RSS to grow continuously until OOM.
\nsys exhibits similar trace-buffer inflation.

In summary, \sysname achieves a total overhead of less than 2\% with constant memory overhead when all three observation sources are active, enabling continuous always-on operation in production environments.
\pytorchprofiler and \nsys are unsuitable as always-on observability solutions due to overhead or training-breaking side effects.

\subsection{Data Volume and Compression}
\label{sec:eval-compression}

This experiment evaluates the per-rank per-step data volume generated under each profiling configuration, and the compression effectiveness of the Processor pipeline, including KDE-based kernel clustering component (\cref{sec:design:compression}).

\begin{table}[t]
\centering
\caption{Per-rank per-step data volume at each processing stage.}
\label{tab:data-volume}
\resizebox{\columnwidth}{!}{%
\begin{tabular}{@{}lrrr@{}}
\toprule
\textbf{Source} & \textbf{Raw} & \textbf{Perfetto trace} & \textbf{Metric Storage} \\
\midrule
CPU call stack (py-spy)     & 250\,KB   & 8\,KB    & ---       \\
Kernel execution (CUPTI)    & 10\,MB    & 420\,KB  & 2.7\,KB  \\
Framework semantics & 394\,KB & 15\,KB   & 16\,KB   \\
\midrule
\textbf{\sysname total}     & \textbf{10.6\,MB} & \textbf{443\,KB} & \textbf{18.7\,KB} \\
\midrule
\nsys (baseline)             & 63.56\,MB & ---      & ---       \\
\pytorchprofiler (baseline)  & 48.76\,MB & ---      & ---       \\
\bottomrule
\end{tabular}%
}
\end{table}

\Cref{tab:data-volume} summarizes the data volume at each processing stage.
\sysname's three observation sources collectively produce approximately 10.6\,MB of raw data per rank per step.
After the Processor converts this into Perfetto-format traces, the volume compresses to 443\,KB.
The structured summaries uploaded to Metric Storage total only 18.7\,KB, of which kernel data achieves a compression ratio of approximately 3{,}700$\times$ through clustering (from 10\,MB to 2.7\,KB).

At 10{,}000-GPU scale with approximately 15 steps per minute, the aggregate upload rate is $10{,}000 \times 15 \times 18.7\,\text{KB} \approx 2.7\,\text{GB/min}$, within time-series database capacity and enabling real-time cross-rank analysis.
In contrast, \nsys and \pytorchprofiler generate 63.56\,MB and 48.76\,MB per rank per step respectively.
At 10{,}000-GPU scale, a single step produces over 600\,GB (\nsys) or 470\,GB (\pytorchprofiler) of trace data, far exceeding the capacity of any online analysis system, and thus only suitable for offline post-hoc analysis.

\section{Case Studies}
\label{sec:eval-casestudy}

This section presents five production case studies to demonstrate the practical effectiveness of \sysname's progressive diagnostic framework. The jobs use Megatron-LM-style hybrid parallelism. We describe each rank by its active parallel coordinates, including data parallelism (DP), tensor parallelism (TP), pipeline parallelism (PP), and expert parallelism (EP); unused dimensions are omitted. The cases cover compute straggler nodes (Case~1), communication link degradation (Case~2), pipeline bubble amplification (Case~3), operator JIT compilation (Case~4), and compute straggler with misleading out-of-band metrics (Case~5).

\subsection{Case 1: Compute Straggler Localization}

This case occurred in a ${\sim}$4{,}000-GPU VLM training job with TP=2 and EP=8. L1 detected a persistent iteration-time regression: step time increased from about 4\,s to more than 200\,s for multiple consecutive steps.  Concurrently, L2 performed CV analysis on semantics events (\texttt{self\_attention}, \texttt{moe\_experts}, \etc) across ranks within the DP group, detecting that the compute-class semantic events on ranks at DP=656 and DP=657 deviated significantly from the group mean, marking them as stragglers.

\begin{figure}[t]
\centering
\begin{subfigure}[t]{\columnwidth}
  \centering
  \includegraphics[width=\textwidth]{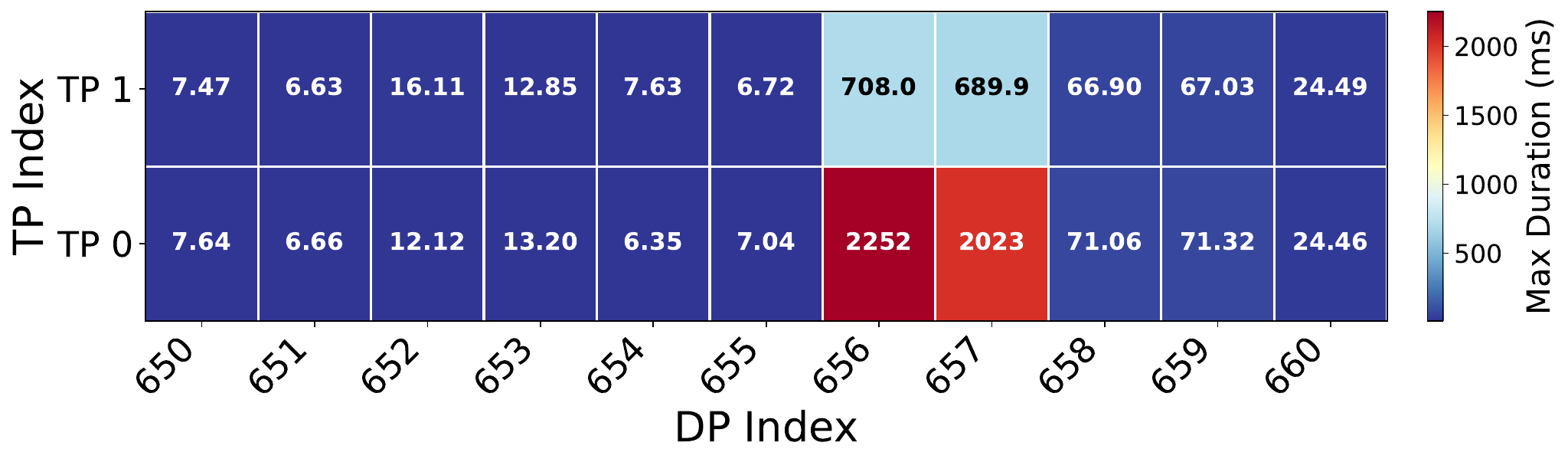}
  \vspace{-0.7cm}
  \caption{\texttt{self\_attention}}
  \label{fig:case1-heatmap-attn}
\end{subfigure}

\begin{subfigure}[b]{\columnwidth}
  \includegraphics[width=\textwidth]{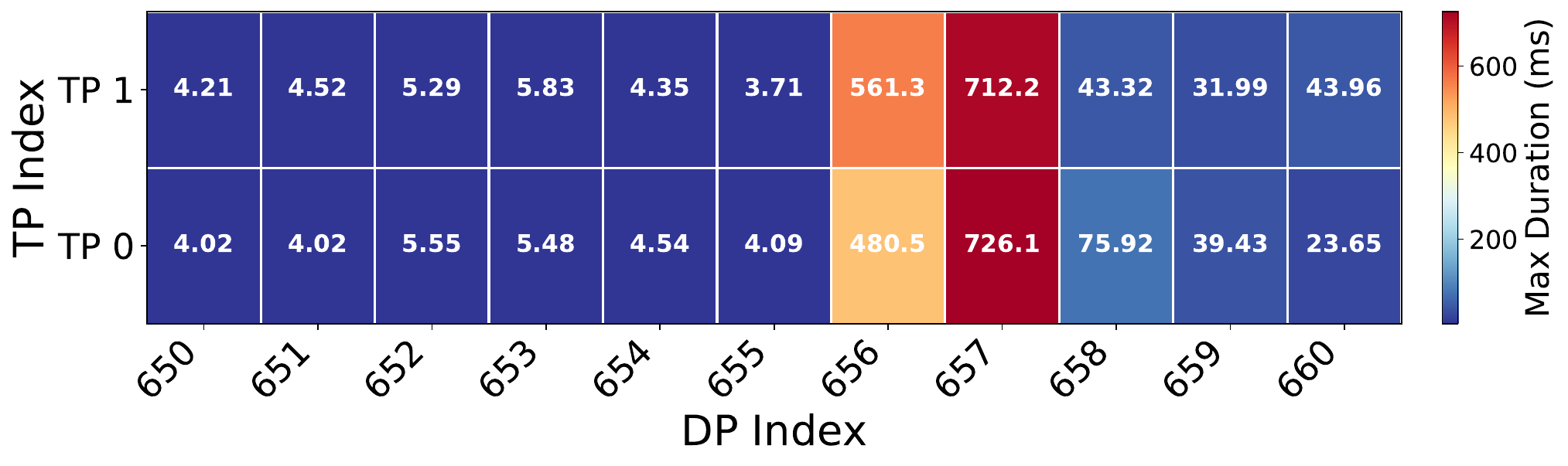}
  \vspace{-0.7cm}
  \caption{\texttt{mlp} / \texttt{grouped\_mlp}}
  \label{fig:case1-heatmap-mlp}
\end{subfigure}
\caption{Case~1: Grafana heatmap of per-rank maximum operator duration. The x-axis is the DP replica index, and the y-axis is the TP index. DP replicas 656 and 657 are outliers across both TP indices, with more than 150$\times$ degradation on compute-only operators.}
\label{fig:case1-heatmap}
\end{figure}

As shown in \cref{fig:case1-heatmap}, the Grafana dashboard visualizes the per-rank maximum operator duration as a heatmap, with the DP replica index on the x-axis and the TP index on the y-axis. Normal ranks spend 6--16\,ms in \texttt{self\_attention} and 4--6\,ms in \texttt{mlp}. In contrast, both TP indices in DP replicas 656 and 657 are outliers: \texttt{self\_attention} reaches 2{,}252\,ms and 2{,}022\,ms, while \texttt{mlp} reaches 480--726\,ms. These phases are compute-only and involve no collective communication, so the anomaly points to local GPU compute degradation rather than synchronization delay. After excluding the affected nodes, training returned to its normal speed.

\vspace{-0.2cm}
\subsection{Case 2: Communication Link Degradation}

This case occurred in a ${\sim}$500-GPU audio-model training job with EP=8. The iteration time was stable with no jitter or regression, but overall throughput remained persistently below the expected baseline. Neither L1 nor L2 detected any anomaly. L3 performed cross-rank $W_1$ comparison on the duration distributions of typical kernels on each rank, identifying significant cross-group distribution shifts on three communication kernel types and triggering an alert.

\begin{figure}[t]
\centering
\begin{subfigure}[t]{0.32\columnwidth}
  \includegraphics[width=\linewidth]{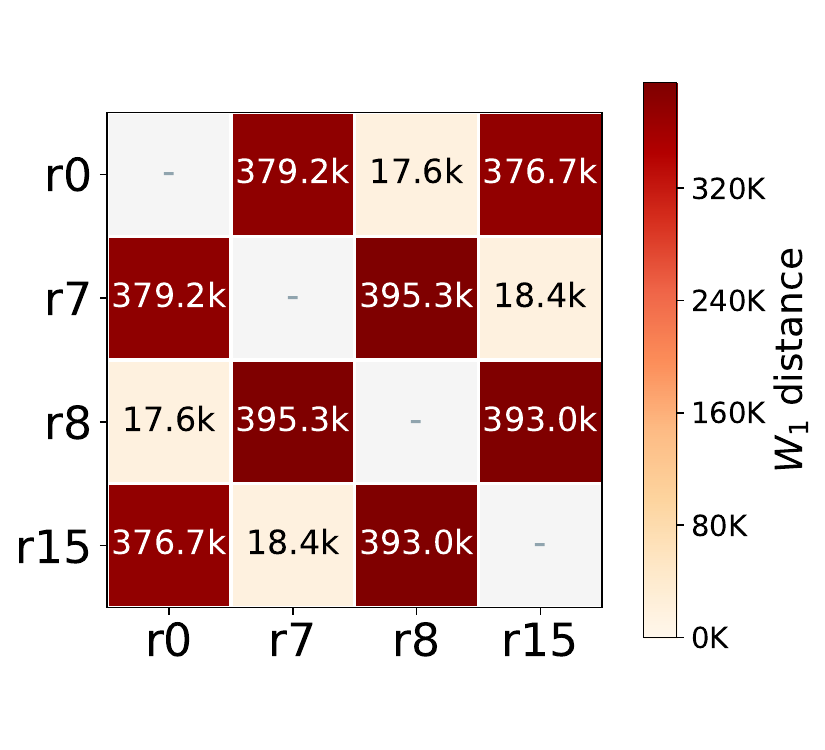}
  \vspace{-0.7cm}
  \caption{\texttt{AllReduce}}
  \label{fig:case2-w1-allreduce}
\end{subfigure}\hfil
\begin{subfigure}[t]{0.32\columnwidth}
  \includegraphics[width=\linewidth]{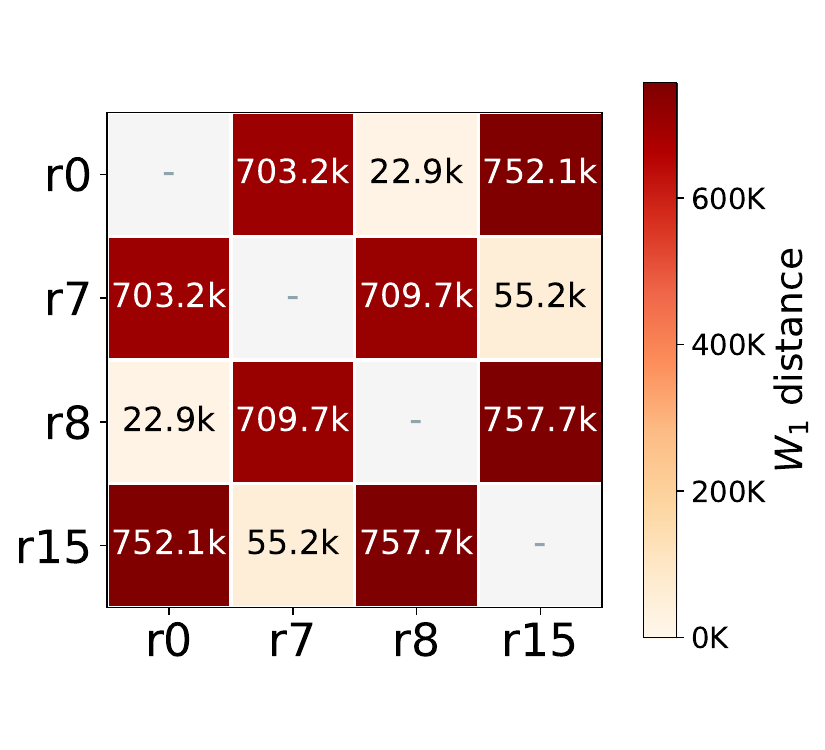}
  \vspace{-0.7cm}
  \caption{\texttt{AllGather}}
  \label{fig:case2-w1-allgather}
\end{subfigure}\hfil
\begin{subfigure}[t]{0.32\columnwidth}
  \includegraphics[width=\linewidth]{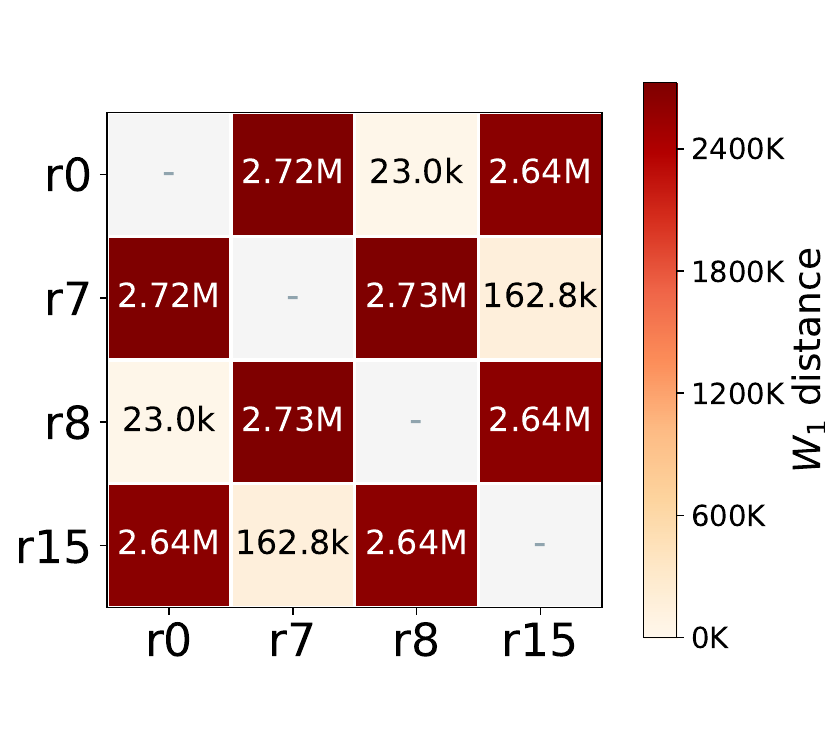}
  \vspace{-0.7cm}
  \caption{\texttt{ReduceScatter}}
  \label{fig:case2-w1-reducescatter}
\end{subfigure}
\caption{Case~2: $W_1$ distance matrices for three communication kernels. Ranks~0 and~8 belong to one EDP group; ranks~7 and~15 belong to another. Intra-group distances are small (17--23k), while inter-group distances are orders of magnitude larger (376k--2.73M), revealing systematic communication degradation in the EDP group containing ranks~7 and~15.}
\label{fig:case2-w1}
\end{figure}

\Cref{fig:case2-w1} compares the pairwise $W_1$ distance matrix of four representative ranks from two expert data parallel (EDP) groups, with ranks~0 and~8 from one group and ranks~7 and~15 from another.
The matrix exhibits a clear grouping pattern: intra-group $W_1$ distances are small (AllReduce: r0--r8\,=\,17.6k, r7--r15\,=\,18.4k), whereas cross-group distances are consistently larger (AllReduce: 376k--395k; AllGather: 703k--757k; ReduceScatter: 2.64M--2.73M). This indicates a systematic difference in communication kernel duration distributions between the EDP group containing ranks~7 and~15 and the group containing ranks~0 and~8.

\begin{figure}[t]
\centering
\includegraphics[width=\columnwidth,page=1]{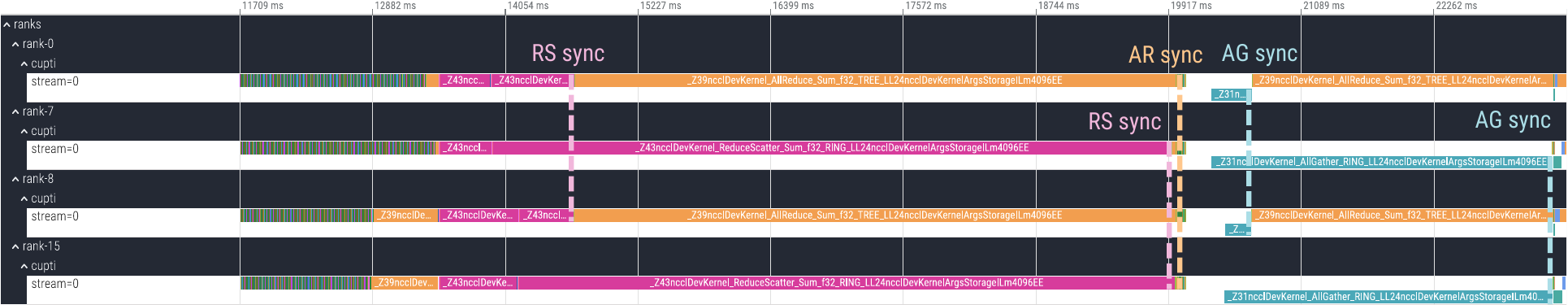}
\caption{Case~2: L4 Perfetto trace of communication kernels. Rank~7 shows longer EDP-internal \texttt{ReduceScatter} and \texttt{AllGather} operations, illustrating network degradation in its own EDP group.}
\label{fig:case2-timeline}
\end{figure}

We further verified this through the L4 Perfetto trace, as shown in \cref{fig:case2-timeline}. The timeline reveals that rank~7's EDP group executes \texttt{AllGather} and \texttt{ReduceScatter} (EDP-internal communication, synchronized only within the EDP group but not across EDP groups) significantly slower than other EDP groups, with no observable waiting time, indicating degradation originates from its own communication link rather than being slowed by other ranks. Conversely, other EDP groups (\eg, rank~0's group) show prolonged \texttt{AllReduce}, because \texttt{AllReduce} must synchronize all EDP groups, and rank~7's EDP group communication latency forces other groups to wait passively. Based on this analysis, we executed a targeted NCCL test on the nodes in the EDP group containing ranks~7 and~15, confirming that two machines in this group had PCIe link hardware faults. After repair, throughput recovered to the expected level. This silent degradation is undetectable by systems relying on iteration time thresholds or heartbeats (\eg, Greyhound~\cite{greyhound}, C4~\cite{c4}), because their detection logic depends on visible performance fluctuations as a trigger condition.

\subsection{Case 3: Pipeline Bubble Amplification}

This case occurred in a ${\sim}$4{,}000-GPU VLM job with TP=4, PP=4, and EP=8.
The job ran slower than expected, yet none of L1, L2, or L3 triggered any automatic alert. We manually inspected per-rank semantics event durations through the Grafana Dashboard, observing that rank~3760's \texttt{backward-compute} median duration was approximately 173\,ms, while other ranks at the same PP index showed only about 92\,ms --- a ratio of about 1.9$\times$. This prompted further investigation of rank~3760's PP group.

\begin{figure}[t]
\centering
\includegraphics[width=\columnwidth,page=2]{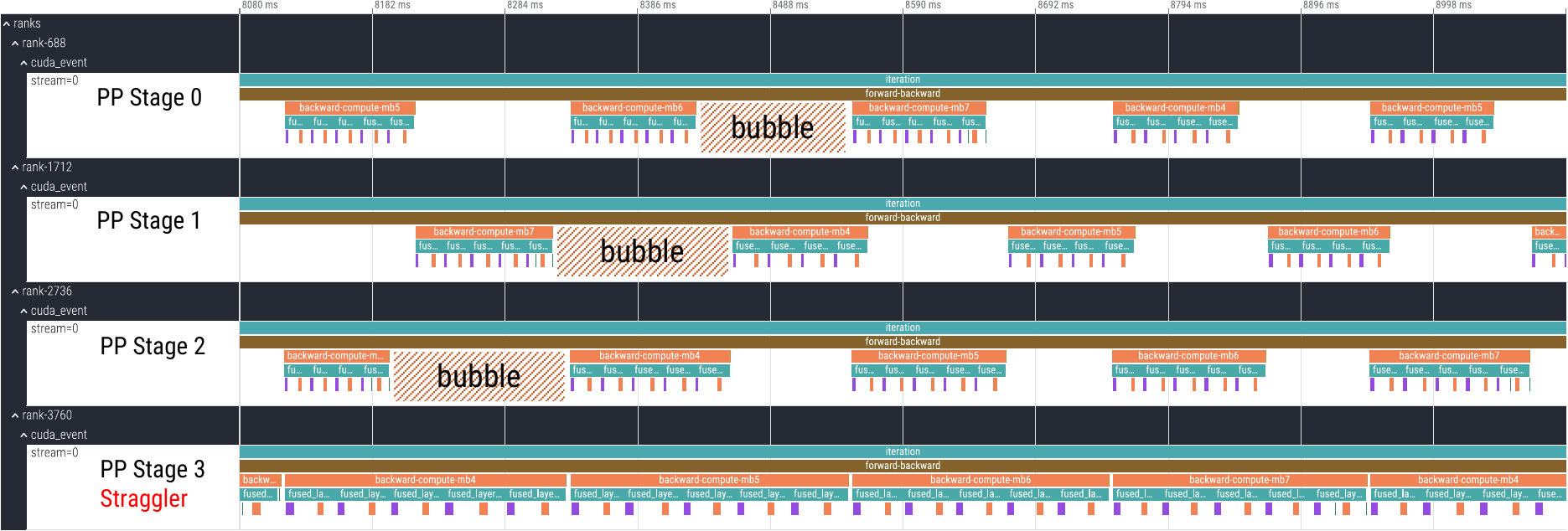}
\caption{Case~3 (a): L4 Perfetto trace of one PP group (ranks~688/1712/2736/3760, PP stages~0--3). The last PP stage, rank~3760, has tightly packed backward-compute events and small bubbles. Upstream stages wait for gradients from the slow downstream stage.}
\label{fig:case3-bubble}
\end{figure}

We retrieved the L4 Perfetto trace and compared the 4~PP stages (ranks~688/1712/2736/3760) within the same PP group, as shown in \cref{fig:case3-bubble}. The timeline exhibits a clear ``asymmetric bubble'' pattern: rank~3760 (the last PP stage) has tightly packed backward-compute events for each micro-batch, with average inter-event bubbles of only 160\,ms; ranks~688/1712/2736 (the first three PP stages) show average bubbles of 227--233\,ms. This indicates that rank~3760 is the compute bottleneck: its GPU runs at full load, filling bubbles with computation, while upstream PP stages wait for downstream gradient propagation, producing idle gaps.

\begin{figure}[t]
\centering
\includegraphics[width=\columnwidth,page=3]{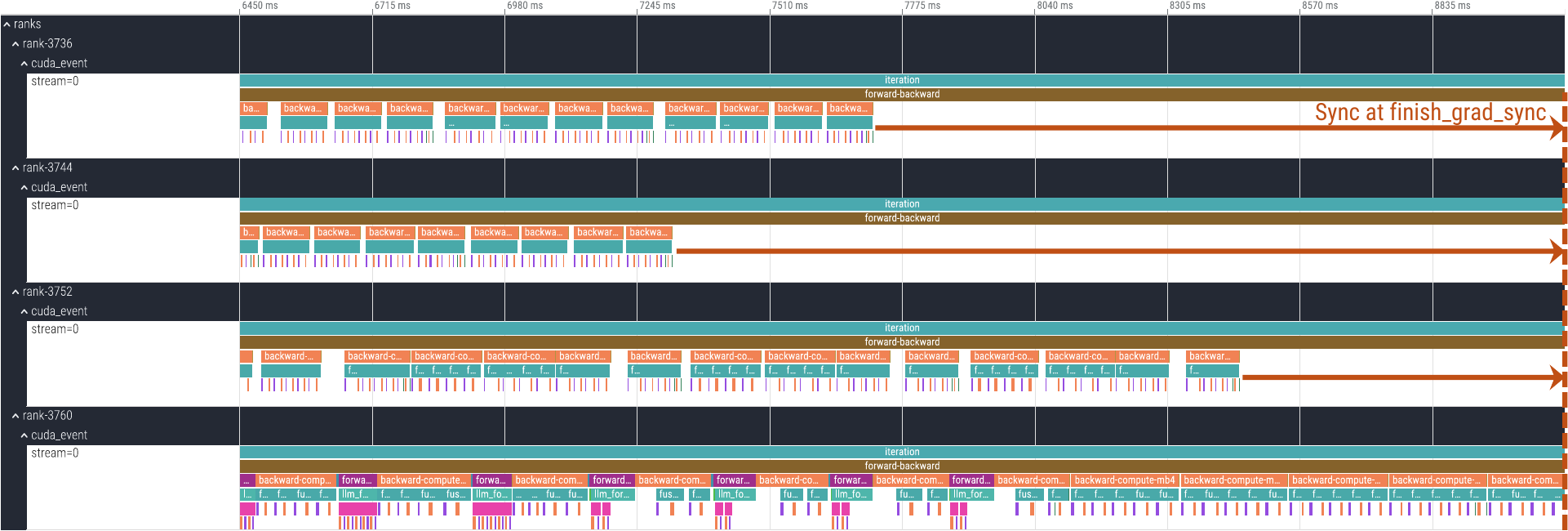}
\caption{Case~3 (b): L4 Perfetto trace of ranks at the same PP index (PP stage~3) across different PP groups (ranks~3736/3744/3752/3760). Rank~3760's backward-compute is denser; yet all ranks' \texttt{forward-backward} durations are aligned at ${\sim}$11{,}178\,ms due to \texttt{finish\_grad\_sync}.}

\label{fig:case3-longtail}
\end{figure}

We further compared ranks at the same PP index (PP stage~3) across different PP groups (ranks~3736/3744/3752/\allowbreak 3760), as shown in \cref{fig:case3-longtail}. The trace shows that rank~3760's backward-compute events consistently finish later than its peers at the same PP stage, confirming it as the sole compute bottleneck. However, all ranks' \texttt{forward\allowbreak{}-backward} total durations are nearly perfectly aligned (approximately 11{,}178\,ms), because the trailing \texttt{finish\_grad\_sync} is a synchronous operation that forces all gradient aggregation to complete. This alignment effect causes the iteration time observed by L1/L2 to show minimal inter-rank differences, effectively masking the bottleneck. After removing the corresponding straggler node, training speed recovered to expectations. Existing systems (\eg, Holmes~\cite{holmes}, Minder~\cite{minder}) operate at iteration/operator or machine-level symptoms and do not model PP pipeline causal dependencies, making it difficult for them to identify straggler nodes concealed by synchronous alignment.

This case reveals two masking mechanisms for straggler nodes under PP parallelism~\cite{gpipe, pipedream, straggler_whatif}. First, pipeline dependency-induced ``bubble transfer'' causes the slow rank's compute bottleneck to propagate through PP dependencies into idle gaps on other ranks, diffusing the anomaly across ranks rather than concentrating it at its source. Second, the alignment effect of \texttt{grad\_sync} forces iteration time to converge across ranks through global synchronization, masking the performance difference. L1, L2, and L3 did not trigger alerts because the anomalous rank's \texttt{backward-compute} peak duration is about 1.9$\times$ that of normal ranks, while natural variation due to different image shapes in VLM tasks can reach 2--3$\times$, and synchronous alignment further flattens the iteration time differences, preventing statistical tests from declaring the deviation significant.

\subsection{Case 4: FlashAttention JIT Compilation}

This case occurred in a ${\sim}$4{,}000-GPU VLM job with TP=4, PP=4, and EP=8. Training exhibited frequent iteration time spikes, with some steps experiencing latency surges of tens of times. L1 detected jitter and triggered an alert. However, L2 and L3 did not identify persistent anomalies because the spikes occurred only occasionally and recovered after a few steps, being diluted within their statistical windows.

\begin{figure}[t]
\centering
\includegraphics[width=\columnwidth,page=4]{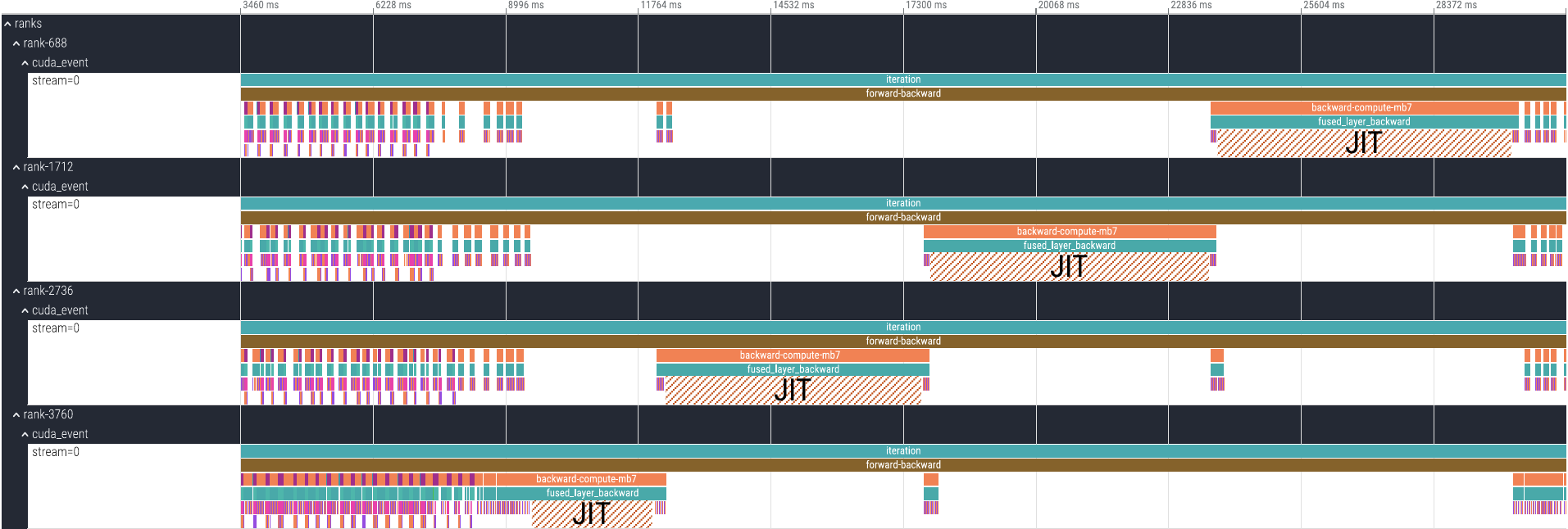}
\caption{Case~4: L4 Perfetto trace of an anomalous step. In the PP group containing rank~688, \texttt{backward-compute-mb7} becomes about 40$\times$ longer than normal. Sparse kernel launches indicate host-side blocking rather than GPU computation.}
\label{fig:case4-jit}
\end{figure}

Through L4 analysis in the Grafana dashboard, we observed a rare but extreme spike: the maximum \texttt{backward-compute} duration in the PP group containing rank~688 was about 40$\times$ larger than that of other PP groups, although the spike was too short-lived to form a persistent L2 anomaly.  We therefore retrieved the Perfetto trace for the anomalous step on rank~688, as shown in \cref{fig:case4-jit}, and compared ranks~688/1712/2736/3760 within the PP group. Within the anomalous step, most \texttt{backward-compute-mb} events take approximately 95--160\,ms, but \texttt{backward-compute-mb7} takes 6{,}303\,ms (approximately 40$\times$ inflation). The internal structure of this anomalous event shows only a single \texttt{fused\_layer\_backward} sub-event executing, with an extremely sparse kernel launch pattern where most time is consumed in host-side waiting rather than GPU computation. Combined with CUDA JIT compilation-related messages observed in the training logs, we confirmed the blocking originated from FlashAttention's JIT compilation.

The recent FlashAttention implementation inherits the IO-aware exact-attention design~\cite{flashattention}, while its CuTe DSL backend~\cite{cutlass_cute_dsl} uses runtime JIT: the first invocation of an uncached kernel configuration lowers DSL code to PTX/cubin and caches the result~\cite{flashattention4}. Under the default configuration, compilation results are cached only in process memory; any fault-tolerance event (fatal node replacement, StepHang recovery, configuration hot-update) that restarts the process clears the cache. Since this job is a VLM with diverse input image shapes, JIT compilation is triggered frequently. The compilation-induced blocking propagates through PP dependencies: when compilation occurs on one rank, downstream PP ranks cannot receive activations or gradients, cascading the delay across the PP group.

The optimization measures adopted include: enabling disk caching via an environment variable so that compilation artifacts persist across process restarts; and pre-enumerating possible shape combinations during a warm-up phase before training to complete compilation ahead of time. After these optimizations, JIT compilation spikes were completely eliminated. For such intermittent anomalies, traditional sampling-based profiling is unable to capture transient events. Even systems with always-on capability cannot trace an iteration-time spike to the specific operator-level blocking source without fine-grained execution traces that preserve per-event timing within each step.

\subsection{Case 5: Compute Straggler with Misleading Out-of-band Metrics}

This case occurred in a ${\sim}$10{,}000-GPU MoE training job with TP=1, PP=9, and EP=32. L1 detected iteration time regression from approximately 30\,s to over 90\,s, triggering an alert. L2 performed CV analysis on semantics events across ranks within the DP group. The compute-only \texttt{mlp} phase showed a CV above 0.8 on ranks~10352--10359 (PP=7, DP=272--279), whose durations were about 5.7$\times$ the group mean.

\begin{figure}[t]
\centering
\begin{subfigure}[b]{\columnwidth}
  \includegraphics[width=\textwidth]{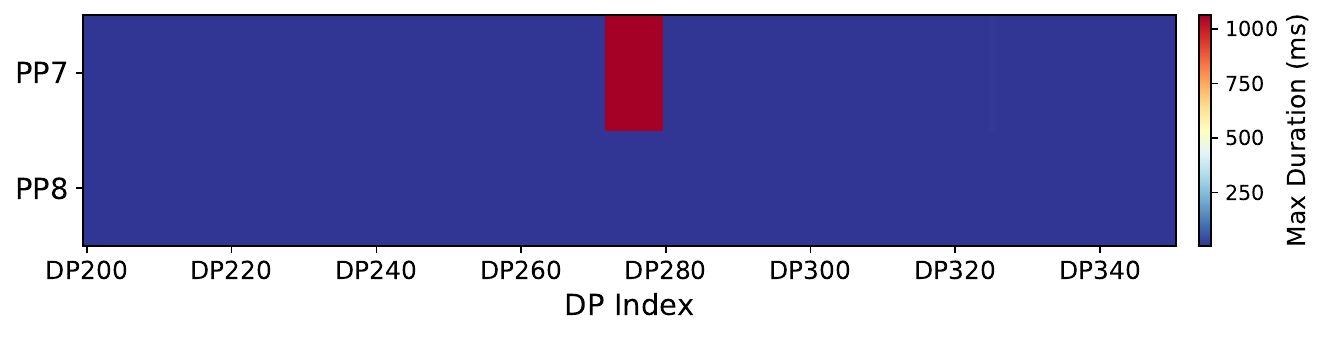}
  \vspace{-0.7cm}
  \caption{\texttt{mlp} / \texttt{grouped\_mlp}}
  \label{fig:case5-heatmap-mlp}
\end{subfigure}

\begin{subfigure}[b]{\columnwidth}
  \includegraphics[width=\textwidth]{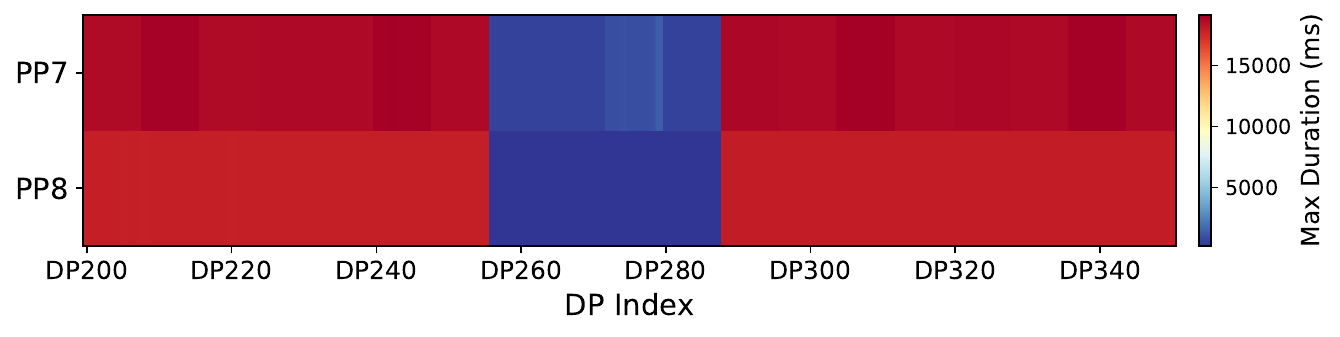}
  \vspace{-0.7cm}
  \caption{\texttt{ReduceScatter}}
  \label{fig:case5-heatmap-rs}
\end{subfigure}
\caption{Case~5: Heatmap of per-rank max duration. The x-axis is the DP replica index, and the y-axis is the PP stage index. (a)~MLP shows extreme degradation at PP=7, DP=272--279 (ranks~10352--10359, ${\sim}$5.7$\times$). (b)~The affected EP group spans DP replicas~256--287 and shows shorter \texttt{ReduceScatter} durations because compute stragglers delay its entry into DP-level communication.}
\label{fig:case5-heatmap}
\end{figure}

As shown in \cref{fig:case5-heatmap}, the Grafana heatmap presents per-rank maximum operator duration with DP index on the x-axis and PP index on the y-axis. In \cref{fig:case5-heatmap-mlp}, most ranks show stable \texttt{mlp} latency around 30--50\,ms, while ranks~10352--10359 (PP=7, DP=272--279) reach 170--280\,ms. \Cref{fig:case5-heatmap-rs} reveals a complementary \emph{inverse} pattern for \texttt{ReduceScatter}: looking vertically, \emph{all} PP stages show \emph{shorter} \texttt{ReduceScatter} durations than surrounding ranks. The 8~slow-compute ranks (10352--10359) reside in the same EP group as 24~other ranks (EP group: ranks~10336--10367, 32~ranks total). The EP-internal dispatch operation requires all 32~ranks to participate; the 8~slow ranks stall this collective, causing the entire EP group to enter the DP-level \texttt{ReduceScatter} late. As shown in the PP=7 row of \cref{fig:case5-heatmap-rs}, because the other EP groups in the same DP group are already ready when this EP group arrives, its \texttt{ReduceScatter} duration appears shorter. Furthermore, PP inter-stage dependencies propagate the delay to all other PP stages in the same PP group: each stage must wait for the slow PP=7 stage to complete before proceeding, causing ranks across all PP stages at DP=256--287 (the full EP group's DP range) to also enter their EP-group \texttt{ReduceScatter} late, exhibiting similarly shorter durations. This confirms compute degradation on ranks~10352--10359 as the root cause.

However, the out-of-band monitoring system reported a ``server port down'' event on one of the affected nodes, which operations initially attributed as the root cause and scheduled for network repair. \sysname's fine-grained tracing, in contrast, clearly showed that the anomaly manifested exclusively on pure-compute operators with no communication involvement, contradicting the network attribution. After replacing the affected nodes, training speed recovered immediately. This case demonstrates a key limitation of out-of-band metrics: communication anomalies observed at the infrastructure level do not necessarily originate from network hardware itself---they can be secondary effects of compute degradation propagating through collective dependencies. Without fine-grained visibility, operators risk misdiagnosing the problem and applying ineffective remediation.

Overall, fail-slow modes in such 10{,}000+ GPU training clusters are highly diverse, and no single diagnostic level can cover all scenarios. \sysname follows a progressive workflow: L1--L3 automatically narrow the scope, while L4/L5 provide on-demand root-cause confirmation, completing the loop from anomaly discovery to localization.

\vspace{0pt}
\section{Discussion}
\label{sec:discussion}

\para{Agent-based intelligent diagnosis.}
The current framework achieves automated anomaly detection and preliminary attribution, but the complete closed loop from detection to root-cause confirmation still relies on engineer expertise. We are exploring integrating LLM agents that take \sysname's detection results as input, combine them with parallel topology and historical fault patterns, and automatically perform multi-round reasoning and tool invocations to produce structured diagnostic reports. Preliminary experience shows that agents can compress average diagnosis time from tens of minutes to the order of minutes, reducing the reliance on manual on-call intervention.

\para{Low invasiveness and generalization.}
Among the three observation mechanisms, kernel execution tracing is injected via environment variables, and CPU call-stack profiling operates through external process sampling; neither requires any modification to training code. Framework semantics instrumentation only adds minimal instrumentation at critical paths. This low invasiveness makes \sysname applicable beyond pre-training. In practice, \sysname has already been extended to reinforcement learning training, and we plan to generalize it to inference serving in the future.

\section{Conclusion}
\label{sec:conclusion}

This paper presents \sysname{}, a real-time, low-overhead, fine-grained tracing and analysis system for 10,000-GPU scale training clusters. \sysname{} decomposes observation into three independently optimized mechanisms, achieving always-on tracing with less than 2\% overhead. The unified data pipeline handles heterogeneous trace ingestion, tiered storage, and online statistical compression, transforming voluminous raw traces into KB-scale structured summaries for real-time cross-rank analysis while persisting complete traces for deep-dive inspection. On top of this, the progressive diagnosis framework narrows the diagnostic scope from tens of thousands of ranks to single-digit suspects. \sysname{} has been deployed on a 10,000+ GPU production cluster for over six months, running stably alongside production training and playing a key role in rapid fail-slow detection and performance optimization.

\bibliographystyle{ACM-Reference-Format}
\bibliography{bib/reference}

\newpage
\appendix

\section{CUPTI Engineering Optimizations}
\label{sec:appendix:cupti}

Beyond the three-path architectural design described in \S\ref{sec:design:cupti}, \sysname introduces three additional engineering optimizations to further reduce overhead and improve stability.

\para{Selective injection targeting goal processes.}
In real training environments, injection-based profiling faces a practical problem: environment variables and runtime contexts are often inherited by auxiliary processes. For example, compilation workers, build toolchains, launcher processes, and child processes spawned by multiprocessing may formally satisfy the conditions for injection, but they are not the actual target workloads that need to be profiled. Indiscriminately enabling CUPTI tracing on these processes not only introduces additional system overhead but also produces large volumes of noise data unrelated to the training workload. Therefore, \sysname introduces a selective injection mechanism: the system initializes the tracing runtime only when the current process matches the target training workload, such as possessing a distributed worker identity and matching command-line characteristics, skipping auxiliary processes such as compilation workers and launchers.

\para{Pre-allocated buffer reuse.}
In CUPTI Activity API-based tracing, buffer management itself can become a significant source of additional overhead. Activity data exhibits pronounced high-frequency and bursty characteristics. If memory is allocated and freed in the callback path each time, it not only introduces additional heap allocation overhead but may also cause allocator lock contention, memory fragmentation, and latency jitter. Therefore, \sysname employs a pre-allocated buffer reuse strategy: the system allocates a fixed number of fixed-size host-side trace buffers once during initialization and reuses them cyclically: CUPTI writes activity records into an idle buffer from the pool; once full, the buffer is handed to the backend for parsing; after parsing completes, the buffer is returned to the pool for reuse. This approach moves memory allocation out of the high-frequency hot path, leaving only lightweight ``take buffer'' and ``return buffer'' operations during online collection.

\para{Bounded resources and backpressure.}
Another common problem in high-frequency tracing systems is that once the backend processing speed cannot keep up with data production, unprocessed data accumulates in memory, eventually causing profiling itself to evolve into a new performance bottleneck. Therefore, \sysname adopts a bounded resources and backpressure control strategy: critical resources in the system, including the trace buffer pool and the backend export queue, are all constrained within explicit budgets. When the backend processing speed cannot keep pace with the frontend collection rate, the system does not unconditionally expand capacity or queue without limit; instead, it controls the additional cost of profiling within acceptable bounds by explicitly dropping partial data and continuously monitoring buffer state. In practice, buffer sizes are determined based on production workload characteristics, and no data dropping has been observed in deployment.

\section{Diagnosis Algorithm Details (L1/L2)}
\label{sec:appendix:diagnosis}

\para{L1: Sliding-window ratio-gated jitter detection.}
This algorithm identifies time periods with significant fluctuations in the iteration time series, executing in two phases. In the first phase (\emph{sensitivity gating}), a sliding window of width $W$ slides over the time series point by point; for each window position, the ratio $r = \max / \min$ of the values within the window is computed. When $r$ exceeds threshold $\theta$, that window is marked as a candidate anomalous region, and adjacent or overlapping candidate regions are merged into contiguous anomalous intervals. In the second phase (\emph{effective width measurement}), for each merged anomalous interval, the median of all points outside the interval is computed as the baseline $b$; within the interval, the longest contiguous sub-segment where all points significantly exceed the baseline is identified, and the width of this longest contiguous sub-segment is the effective jitter width. The two-phase design resolves the inherent ``smearing'' effect of sliding windows: when a narrow spike much smaller than $W$ appears, the first phase inevitably expands the candidate region to at least width $W$. The second phase then precisely recovers the true anomalous time span through baseline exceedance measurement.

\para{L1: Full-scan change-point detection for regression.}
This algorithm searches for the most significant single change point in the iteration time series. The algorithm iterates through every valid split point $t$ in the sequence, computing the mean $\mu_L$, $\mu_R$ and relative standard deviation $\sigma_L / \mu_L$, $\sigma_R / \mu_R$ for the two segments before and after the split. A valid change point must simultaneously satisfy: the regression ratio $\mu_R / \mu_L$ exceeds a minimum threshold (ensuring practically meaningful magnitude), and the relative standard deviation on both sides falls below an upper limit (ensuring both segments are internally stable). The valid split point with the largest regression ratio is selected as the detection result.

\para{L2: CV and z-score computation.}
Let the average duration of a given event on each rank within parallelism group $G$ during the anomalous time window be $\{\bar{x}_r\}_{r \in G}$. \sysname computes the group mean and standard deviation:
\begin{equation}
\setlength{\abovedisplayskip}{1pt}
\setlength{\belowdisplayskip}{1pt}
\mu_G = \frac{1}{|G|} \sum_{r \in G} \bar{x}_r, \quad \sigma_G = \sqrt{\frac{1}{|G|-1} \sum_{r \in G} (\bar{x}_r - \mu_G)^2}
\end{equation}
The coefficient of variation $\text{CV} = \sigma_G / \mu_G$ quantifies the degree of intra-group inconsistency, classified into three levels: balanced ($\text{CV} < 0.02$), mild imbalance ($0.02 \leq \text{CV} < 0.05$), and severe imbalance ($\text{CV} \geq 0.05$). Each rank's z-score $z_r = (\bar{x}_r - \mu_G) / \sigma_G$ identifies stragglers when exceeding the threshold.

\section{Experimental Setup}
\label{sec:appendix:setup}

\begin{table}[t]
\centering
\caption{Experimental model configuration.}
\label{tab:model-config-appendix}
\resizebox{0.85\columnwidth}{!}{%
\begin{tabular}{@{}ll@{}}
\toprule
\textbf{Parameter} & \textbf{Value} \\
\midrule
Transformer layers      & 36 \\
Hidden size             & 2048 \\
FFN hidden size         & 6912 \\
Attention heads / KV groups & 32 / 4 \\
Number of experts / Top-K   & 128 / 8 \\
Expert FFN hidden size  & 768 \\
Sequence length         & 4096 \\
Micro-batch / Global-batch  & 1 / 32 \\
Precision               & BF16 \\
Parallelism strategy    & TP=1, PP=1, EP=8 \\
Optimizer               & Distributed Adam (ZeRO-1) \\
\bottomrule
\end{tabular}%
}
\end{table}

Experiments are conducted on an 8-GPU node with intra-node GPUs interconnected via NVLink. We train the HunYuan-V3 Preview model~\cite{hunyuan}, which is based on a Mixture-of-Experts (MoE) architecture. The primary configuration parameters are summarized in Table~\ref{tab:model-config-appendix}. Under this configuration, each GPU holds $128/8=16$ local experts, and GPUs exchange tokens via AllToAll communication. Training is run with performance optimizations including overlap-grad-reduce, overlap-param-gather, and moe-permute-fusion enabled.

\section{Fault Diagnosis Capability}
\label{sec:appendix:diagnosis-capability}

\begin{table*}[!tb]
\centering
\caption{Summary of \sysname fault diagnosis capabilities. Each fault category is detected through \sysname's progressive diagnostic levels (L1--L3), with L4/L5 available for root-cause confirmation.}
\label{tab:diagnosis-capability}
\small
\begin{tabular}{@{}p{2.8cm}p{3.5cm}cp{3cm}p{4cm}@{}}
\toprule
\textbf{Category} & \textbf{Fault Type} & \textbf{Tier} & \textbf{Symptom} & \textbf{\sysname Localization} \\
\midrule
\multirow{3}{2.8cm}{Compute hardware degradation}
& GPU frequency throttling & L2+L3 & Compute kernels (GEMM, attention) consistently slower on specific rank & Straggler rank identified via CV; anomalous compute kernels flagged at L3 \\
\cmidrule(l){2-5}
& PCIe bandwidth degradation & L3 & Data-transfer kernels (memcpy H2D/D2H) slower on affected rank & Anomaly detected on memory-copy kernels \\
\cmidrule(l){2-5}
& ECC error correction overhead & L1+L3 & Intermittent iteration spikes; tail latency on specific kernels & Jitter detected at L1; tail anomaly on specific kernels at L3 \\
\midrule
\multirow{3}{2.8cm}{Communication degradation}
& NVLink bandwidth degradation & L2+L3 & Intra-node AllReduce/ReduceScatter slower on affected rank & Straggler in EP/DP group; communication kernels flagged at L3 \\
\cmidrule(l){2-5}
& Inter-node network congestion & L1+L2 & Iteration time jitter; communication phase prolonged & Jitter at L1; communication phase identified as bottleneck at L2 \\
\cmidrule(l){2-5}
& RDMA link quality degradation & L3 & AllGather/ReduceScatter kernels show persistent slowdown & Anomaly on specific NCCL kernels on affected stream \\
\midrule
\multirow{3}{2.8cm}{Host-side issues}
& Python GC pause & L1+L5 & Iteration time spikes; GPU execution normal but iteration prolonged & Jitter at L1; L2 shows no straggler in compute phases; CPU stack confirms GC \\
\cmidrule(l){2-5}
& Data loading stall & L2+L5 & GPU idle gap before forward-compute & Semantics shows prolonged gap; CPU stack shows I/O wait \\
\cmidrule(l){2-5}
& CPU contention (co-location) & L1+L2 & Sustained throughput regression on affected node & Regression at L1; all phases of affected rank slow at L2 \\
\midrule
\multirow{2}{2.8cm}{Framework/ configuration}
& MoE expert load imbalance & L2 & \texttt{moe\_experts} duration varies across ranks in EP group & CV analysis on EP group identifies imbalanced ranks \\
\cmidrule(l){2-5}
& Suboptimal comm-compute overlap & L4 & Communication kernels not overlapped with compute in Perfetto trace & Perfetto timeline reveals serialized execution pattern \\
\bottomrule
\end{tabular}
\end{table*}

Table~\ref{tab:diagnosis-capability} summarizes the fault categories, representative symptoms, and corresponding diagnostic pathways through \sysname's progressive framework. The system covers four major categories of fail-slow faults encountered in production: compute hardware degradation (GPU frequency throttling, PCIe bandwidth degradation, ECC error correction), communication degradation (NVLink, inter-node congestion, RDMA link quality), host-side issues (Python GC, data loading, CPU contention), and framework/configuration issues (MoE load imbalance, suboptimal overlap).

\end{document}